\documentclass[pdflatex,sn-nature]{sn-jnl}% Style for submissions to Nature Portfolio journals
\usepackage{graphicx}%
\usepackage{multirow}%
\usepackage{amsmath,amssymb,amsfonts}%
\usepackage{amsthm}%
\usepackage{mathrsfs}%
\usepackage[title]{appendix}%
\usepackage{xcolor}%
\usepackage{textcomp}%
\usepackage{manyfoot}%
\usepackage{booktabs}%
\usepackage{algorithm}%
\usepackage{algorithmicx}%
\usepackage{algpseudocode}%
\usepackage{listings}%
\theoremstyle{thmstyleone}%
\theoremstyle{thmstyletwo}%

\theoremstyle{thmstylethree}%

\begin{document}

\title[Article Title]{Predictable Drifts in Collective Cultural Attention: Evidence from Nation-Level Library Takeout Data}

\author[1,2,3,4]{\fnm{Anders} \sur{Weile}}

\author[1,2]{\fnm{Vedran} \sur{Sekara}}

\affil[1]{\orgdiv{Networks, Data and Society (NERDS)}, \orgname{IT University of Copenhagen}, \orgaddress{\street{Rued Langgaards Vej 7}, \city{Copenhagen S}, \postcode{2300}, \country{Denmark}}}

\affil[2]{\orgname{Pioneer Centre for Artificial Intelligence}, \orgaddress{\street{Øster Voldgade 3}, \city{Copenhagen}, \postcode{1350}, \country{Denmark}}}

\affil[3]{\orgdiv{Center for Social Data Science (SODAS)}, \orgname{University of Copenhagen}, \orgaddress{\street{Øster Farimagsgade 5}, \city{Copenhagen}, \postcode{1353}, \country{Denmark}}}

\affil[4]{\orgdiv{Guest Researcher}, \orgname{Statistics Denmark}, \orgaddress{\street{Sankt Kjelds Plads 11}, \city{Copenhagen}, \postcode{2100}, \country{Denmark}}}

\abstract{Predicting changes in consumer attention for cultural products, such as books, movies, and songs, is notoriously difficult. 
Past research suggests intrinsic limits for predicting consumer attention towards individual products.
However, little is known about the limits for predicting shifts in collective attention.
Here, we analyze five years of nationwide library loan data for almost 3 million individuals, comprising over 136 million loans of more than 750,000 unique titles.
We find that culture, as measured by popularity distributions of loaned books, drifts continually from month to month at a near-constant rate, leading to a growing divergence over time, and that drift varies between book genres.
By linking book loans to registry data, we investigate the influence of age, sex, educational level, and residential area type on cultural drift, finding heterogeneous effects.
Our findings have important implications for market forecasting and algorithmic recommender systems, highlighting the need to account for drift dynamics.}


\maketitle

\newpage 
\noindent Cultural products drive billion-dollar industries, influence public discourse, and shape our personal and collective identities.
Predicting which products will capture the attention of audiences holds significant economic and cultural value.
Yet, it is notoriously difficult to obtain accurate predictions, even despite advances in predictive modeling and the growing availability of large-scale cultural consumption data~\cite{bradlow_bayesian_2001, dellarocas_exploring_2007, asur_predicting_2010, goel_predicting_2010, goel_prediction_2010, mestyan_early_2013, shulman_predictability_2016, yucesoy_success_2018, interiano_musical_2018, wang_success_2019, lee_social_2024}. Seminal work suggests that this difficulty reflects not only data or modeling limitations, but more fundamental limits to prediction in cultural markets, where an abundance of products compete for scarce attention, and consumer choices are strongly shaped by social influence, making outcomes highly contingent and path-dependent~\cite{salganik_experimental_2006, salganik_leading_2008}. These features are not unique to cultural markets but apply to other domains like news and online content where consumption and attention are meaningfully collective phenomena. Across these settings, a small number of products typically captures a disproportionately large share of consumer attention~\cite{cox_concentration_1995, crain_consumer_2002, devany_uncertainty_1999, sorensen_bestseller_2007, newman_power_2005}. Most studies of collective attention focus on the most popular products and major events that capture a large share of our collective attention~\cite{wu_novelty_2007, lehmann_dynamical_2012, lin_rising_2014, candia_universal_2018, Lorenz-spreen_accelerating_2019, dedomenico_unraveling_2020}.

Existing research offers little insight into how collective attention changes across an entire market. Understanding these system-level dynamics is important because they may reveal regularities and constraints that remain unobserved when focusing only on products in isolation. Several important questions remain open. How fast does collective attention change overall? Are changes to collective attention universal across products and consumer groups? Are they predictable? Answers to these questions have implications for our theoretical understanding of cultural market dynamics and for practical applications involving models of market dynamics, such as developing robust recommender systems. Resolving them requires comprehensive cultural consumption data that moves beyond the current focus on hit products as well as an analytical framework for quantifying changes in collective attention at the market level.

Here, we use a unique nation-scale book loan dataset which contains records from all public libraries in Denmark, featuring 136 million book loans from 3 million individuals (almost half of the Danish population), to study how cultural attention evolves at a societal scale.
While libraries are not a traditional economic market per se, as individuals do not have to pay to loan books, books in a library compete in what is often called an `attention market'. In these markets human attention is the limited commodity which these books compete for. We study how attention changes in this system over time, and which factors drive the change. 

In Denmark, library borrowing has a low barrier to entry since library services are well-funded, free to use, and accessible anywhere in the country. In addition, Danish public libraries have a wide coverage, and any material present in the library system can be ordered from any local library. 
For these reasons, libraries serve as a primary means of accessing books for Danish citizens. According to the best available data on book sales in Denmark, the number of book loans from public libraries is consistently almost three times the number of book sales (see Supplementary Fig.~S8 based on~\cite{statisticsdenmark_book_}). In addition, past work has established that loans in our data accurately reflect loaners' self-reported literary tastes, suggesting that library loans is a good indicator of broader book consumption~\cite{blaabaek_how_2025}.
Further, Denmark is one of the most digitalized countries in the world, making it possible to link each individual book loan to registry data, which provides information about library users' demographic characteristics, such as age, sex, education level, zip code, etc.
We use the unprecedented scope of this dataset to identify changes in collective consumer attention for a wide range of books across a large number of consumers.

Our approach differs fundamentally from prior work on collective attention. Rather than focusing on the individual trajectories of popular items in isolation, we analyze the changes in consumer attention across all items in the market, which we refer to as \textit{drift} in collective attention. We borrow the term `drift' from the machine learning literature to describe the shifts in the underlying data distributions that machine learning model are trained on, which can negatively affect model performance~\cite{widmer1996learning,quinonero2008dataset}. Our approach to measuring changes in collective attention eliminates the need to filter away less popular products, capturing attention dynamics across the full popularity spectrum. In addition, it requires no assumptions about how individual items gain or lose consumer attention, accommodating diversity in popularity trajectories.

Our analysis is split up into three parts. First, we quantify drift in collective attention, finding that culture has a constant turnover rate across months with seasonal variation. Second, we dig into which demographic groups contribute to this drift, finding that age and sex have a large influence. Third, we account for the contributions of each unique book title to collective attention drift. We use our findings to build a simple model that can accurately estimate the future magnitude of drift without needing to account for individual product trajectories.

\section*{Results}\label{sec2}
\begin{figure}[!htbp]
\centering
\includegraphics[width=0.9\textwidth]{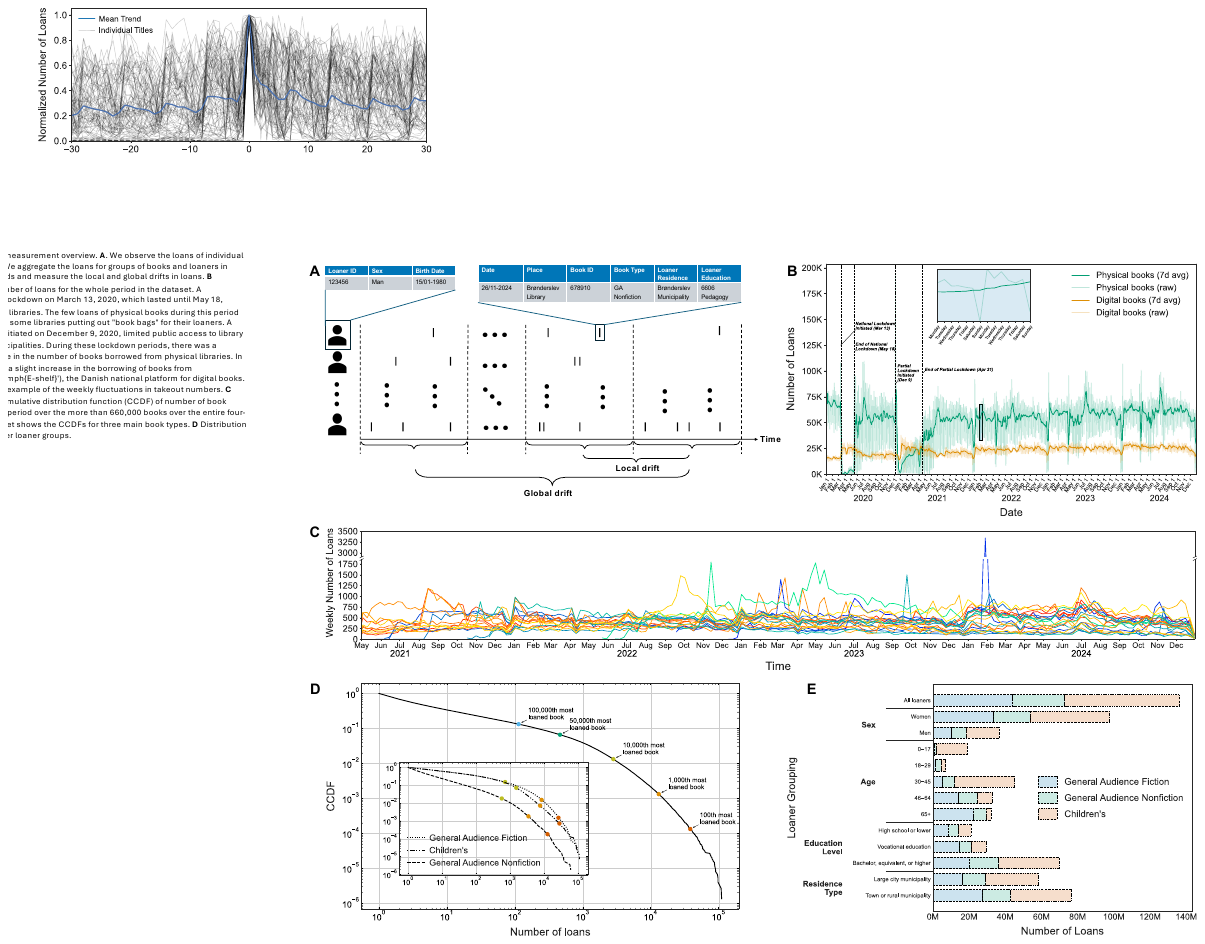}
\caption{\textbf{Dataset and measurement overview.}
\textbf{A,} Each entry in the dataset corresponds to the loan of a single book, illustrated as vertical lines. Some loaner attributes (sex and birthdate) are static, whereas others (loaner residence and education level) are dynamic. We aggregate loans into monthly time periods and measure the local (between successive months) and global (between a given period and a reference month) drift in cultural attention.
\textbf{B,} Aggregate daily number of loans. A nationwide COVID-19 lockdown, effective March 13, 2020, and lasting until May 18, 2020, shut down all libraries. The few loans of physical books during this period can be attributed to some libraries putting out `book bags' for their loaners. A partial lockdown, initiated on December 9, 2020, limited public access to libraries in 38 municipalities. During these lockdown periods, there was a noticeable decrease in the number of books borrowed from physical libraries and an increase in digital loans. The inset shows an example of the weekly fluctuations in takeout numbers over two weeks from 24/01/2022 to 06/02/2022.
\textbf{C,} Weekly number of loans for 20 popular books in the dataset. The y-axis is discontinued because of one book with a very high number of loans around Feb, 2024.
\textbf{D,} Complementary cumulative distribution function (CCDF) of the number of book loans for the more than 750,000 books over the entire five-year period. The inset shows the CCDFs for three main categories of books.
\textbf{E,} Number of loans for different sociodemographic subgroups.}\label{fig_1}
\end{figure}

We use comprehensive records of all Danish public library loans (excluding university and school libraries) for five years from start 2020 to end 2024. Each entry in our dataset corresponds to the loan of a single book and contains information on the date of the loan, a unique identifier of the book, a pseudonymized loaner ID, the library where the book was loaned, as well as metadata on the books and loaners (see Fig.~\ref{fig_1}A and Methods). The dataset contains more than 136 million unique book loans from 513 libraries, including eBooks from digital services (see Supplementary Fig.~S9 for the distribution of loans across material types) and features loans from almost 50\% of the Danish population. Our analysis focuses on three broad book genre categories: general audience fiction, general audience nonfiction, and children's books, which encompasses both children's and young adult literature. A single book title has multiple identifiers for different media (physical book, audiobook, etc.) and editions. For our main results, we treat each variant of a book as a single entity (see Methods for details on how we identify individual titles). We link loaner IDs to national registry data, in a privacy-preserving manner, to obtain sociodemographic information, including age, sex, education level, and area of residence for each loaner (see Methods).

To quantify collective attention drift in library book consumption, we use the Jensen-Shannon Divergence (JSD). The JSD is an information-theoretic measure that is well-suited to measure changes in sequences of frequency distributions~\cite{grosse_analysis_2002}. Within the social science literature, JSD and related measures have mainly been used to analyze changes in text~\cite{dias_using_2018, barron_individuals_2018, murdock_exploration_2017, gallagher_divergent_2018, gallagher_generalized_2021, gerlach_similarity_2016, nielbo_pandemic_2023, pechenick_characterizing_2015}. In this study, we use it on sequences of monthly book popularity distributions (see Methods for technical details). We use months as the aggregation level in our main results since it allows us to capture changes in consumption without being overly sensitive to minor variations in book popularity that do not reflect meaningful changes in consumer attention. A JSD of 0 indicates the popularity distributions are identical between loans in two months; a JSD of 1, means entirely different books have been loaned. Values between 0 and 1 represent a partial reallocation of collective attention.

We distinguish between two main types of drift: \emph{local} drift, which quantifies the JSD between a given month and the preceding month, and \emph{global} drift, which measures drift between a reference month and all subsequent months (Fig.~\ref{fig_1}A). The two drift types reveal distinct patterns in how library book popularity evolves. While local drift captures immediate shifts in book consumption, global drift quantifies long-term changes to loaning behavior. 

Over the 5-year-span of our dataset, we observe variation in the number of library loans (Fig.~\ref{fig_1}B). In particular, two COVID-19-related lockdowns restricted library access in Denmark, resulting in a dramatic drop in how many physical books were loaned as well as an increase in digital book loans. Additionally, citizens tend to loan more books in the Summer holidays and less when libraries are typically closed (on Sundays and in the Winter holidays). To minimize noise and the effects of COVID-19, our analysis focuses on the period after the last lockdown (from May 2021 to December 2024). There is considerable heterogeneity in how the popularity of individual titles evolves over time, reflecting the difficulty of predicting individual trajectories (Fig.~\ref{fig_1}C). While some books experience sharp peaks followed by steady decline, others maintain more stable popularity levels or exhibit complex patterns, including delayed and recurring peaks in popularity.

Collective consumer attention is highly unequally distributed among the over 750,000 titles in our dataset (Fig.~\ref{fig_1}D). An unequal popularity distribution is not unique to our dataset but is a commonly observed pattern across cultural markets~\cite{cox_concentration_1995, crain_consumer_2002, devany_uncertainty_1999, sorensen_bestseller_2007, newman_power_2005}. Loans appear to be most unequally distributed among fiction books targeted a general audience, and most equally distributed for nonfiction books (see inset in Fig.~\ref{fig_1}D). The shape of the popularity distributions is constant across months (see Supplementary Fig.~S10). This means that drift in collective attention between two months capture changes in which books hold which positions in the rank-frequency distribution rather than simply changes in the shape of the distribution, such as increased or decreased diversity in book loans.

Loaner subgroups are not equally represented in this dataset (Fig.~\ref{fig_1}E). For instance, there is a clear gap between the sexes, with women lending more books. Additionally, book types are not evenly distributed among loaner subgroups, especially among different age segments. Unsurprisingly, young loaners (0-17 years) almost exclusively loan children's books. Loaners aged 30 to 45 take out many children's books and relatively few general audience fiction books, whereas 65+ loaners borrow very few children's books and many general audience fiction books. In Supplementary Fig.~S11, we show the distribution of loans across different book types for each subgroup. Further, Supplementary Fig.~S2 demonstrates that these distributions are stable over time.

\subsection*{Patterns of collective drift}
\begin{figure}[!htbp]
\centering
\includegraphics[width=0.9\textwidth]{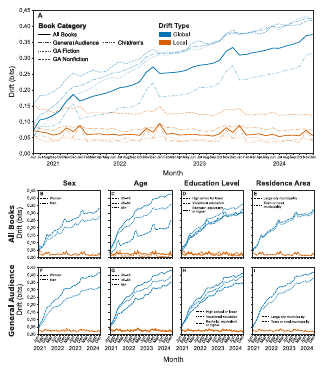}
\caption{\textbf{Local and global drift in cultural attention.}
\textbf{A,} Local (red) and global (blue) drift across all book categories and loaner demographics. We estimate JSD in bits to ensure that drift is bounded between 0 and 1. To pinpoint the contribution to drift from different book categories we use partially overlapping categories. `All books' contains both `general audience' and `children's' categories, and `general audience' (GA) is further split into `fiction' and `nonfiction' categories.
\textbf{B-I,} Local and global drift split up according to population demographics and book category. To minimize the effects of random fluctuations and noise, we focus only on the top 10,000 most popular books. \textbf{B-E,} shows drift for the top 10,000 most popular books, while panels \textbf{F-I} show drift for the top 10,000 most popular general audience books. Shaded areas around lines (barely visible) show $2 \sigma$ bootstrap errors.}\label{fig_2}
\end{figure}

We start our analysis at the most general level by quantifying drift in popularity distributions across all books and loaners.
Fig.~\ref{fig_2}A shows the temporal dynamics of local and global drift. 
Local drift is characterized by being near-constant between successive months, punctuated only by small recurring peaks during Winter periods (November to January; red, solid line Fig.~\ref{fig_2}A).
These deviations stem from season-specific titles, which are predominantly Christmas-themed, becoming popular in November and December (see Supplementary Section 1.3).
Global drift, however, grows continually from the reference month of May 2021 (blue, solid line Fig.~\ref{fig_2}A), and show similar season-specific bumps. Taken together, this suggests two primary mechanisms of drift: a gradual, persistent shift in loaner preferences and a seasonal component causing momentary increases in collective attention. 

To explore whether patterns are similar across different book categories, we compute local and global drift for three broad categories of books, which are available in our dataset: general audience fiction, general audience nonfiction, and children's books. For all book categories, local drifts are stable, meaning that collective attention has a constant rate of drift, while global drift grows over time (dashed and dotted lines in Fig.~\ref{fig_2}A). However, drift patterns differ in notable ways. Specifically, general audience nonfiction books have a larger rate of local drift compared to general audience fiction and children's books. One explanation for this is that a large proportion of nonfiction books are rarely loaned, as seen in Fig~\ref{fig_1}D, inset. When we exclude infrequently loaned books, the local drift rate for nonfiction books becomes less extreme, but it is still higher than drift for fiction and children's books (see Supplementary Fig.~S13). For global drift, we observe different dynamics. Drift in children's books grows at a considerably slower rate than in general audience books.
Lastly, for both local and global drift, the seasonal peaks (around Christmas) are more pronounced for children’s books, less so for general audience nonfiction books, and nearly absent for general audience fiction books.

We assess the robustness of our findings through a series of complementary analyses. The observed patterns are qualitatively robust when using generalized measures of drift (Supplementary Section 1.1 and Supplementary Figs.~S14-S19) and for different temporal aggregation levels other than months (Supplementary Figs.~S19-S22). Our findings also differ substantially from the dynamics produced by a null model in which changes arise solely from random fluctuations in the popularity of books (Supplementary Section 1.2). Another potential concern is that our results could be affected by the limited supply of books in libraries, as we quantify drift from the number of loans rather than the number of requests for a book. Indeed digital books (e-books and audiobooks), which have an unlimited supply, typically have a high number of loans for a shorter time than physical books (Supplementary Fig.~23), suggesting that the supply of physical books cannot always follow the demand. However, drift patterns measured separately for physical and digital books are nearly identical (Supplementary Fig.~24), indicating that supply constraints do not drive our results.

\subsection*{Drift split up according to population subgroups}
Prior work has shown that sociodemographic groups, defined by sex, age, education level, and area of residence to exhibit differences in the diversity of cultural consumption~\cite{park_understanding_2015, anderson_algorithmic_2020}. Yet, to our knowledge, no study has quantified how the collective attention of these segments evolves over time. To compare different subgroups of the loaner population, we account for the fact that drift computed from raw frequencies capture both changes to the underlying collective attention and random fluctuations in the popularity of less popular books. To eliminate the effects of random fluctuations, we compute drift exclusively for the top 10,000 most popular books within each sociodemographic subgroup and correct for any residual bias using bootstrap methods~\cite{dedeo_bootstrap_2013} (see Methods for details). For the comparison of loaners in different age bins, we exclude loaners below the age of 30 as loaners from 0 to 17 almost exclusively lend children's books, and loaners between 18 and 29 have very few loans.

We find substantial differences in drift between men and women, where the latter have a larger global drift and stronger recurring seasonal peaks, while differences in local drift are minuscule (Fig.~\ref{fig_2}B). Additionally, it is worth noting that the frequency of seasonal peaks differs between men and women. 
While women show seasonal peaks centered around Christmas, men, in addition, also have summer peaks. For loaners of different ages we also find large differences in drift (Fig.~\ref{fig_2}C).  Both local and global drift are considerably lower for 30-45-year-old loaners than for older ones, and the seasonal component is stronger. 
For other sociodemographic subdivisions (education level and area of residence), differences in local drift are not pronounced, and we observe only small to moderate differences between global drift over time (Figs.~\ref{fig_2}D-E).

To investigate the role of children's books on drift for individual demographic groups, we compare the results from Fig.~\ref{fig_2}B-E with drift specifically calculated for the 10,000 most popular general audience books (i.e., disregarding loans of children's books). Here, we observe that the differences in global drift between men and women becomes larger when we only consider drift in general audience books (Fig.~\ref{fig_2}F). In addition, disregarding children's books also makes the differences in global drift more prominent for education- and residence area-groups (Fig.~\ref{fig_2}H-I). Here we observe that drift is higher for women, for more educated individuals (Bachelor or higher), and for individuals living in large cities. Differences between age groups, however, get smaller (Fig.~\ref{fig_2}G). Disregarding children's books reduces drift to the same order of magnitude as sex, where differences between older demographics (65+ year-olds) and younger demographics (30-45-year-olds) become similar to that of men and women.
These differences are robust when books are further split up into general audience fiction and general audience nonfiction (Supplementary Fig.~S25). 
When it comes to seasonal peaks, we find that removing children's books eliminates the regularly occurring peaks for both men and women (Fig.~\ref{fig_2}F), and 30-45-year-olds (Fig.~\ref{fig_2}G), consistent with our previous finding that children's books drive the seasonal peaks nationally.

\subsection*{Contributions from individual books}
Until now, we have considered books collectively; here, we untangle the importance of individual items on drift and study what role individual books play in collective attention drift. We do this as much of the focus in both media and previous studies is on big hits that occupy our collective attention at any given moment~\cite{bradlow_bayesian_2001, dellarocas_exploring_2007, asur_predicting_2010, goel_predicting_2010, yucesoy_success_2018, interiano_musical_2018, candia_universal_2018, mestyan_early_2013}. However, it is unclear how important individual items are for explaining overall drift in collective attention. To investigate this, we measure how much each book contributes to (nation-scale) local and global drift. For each month, we divide all books into five broad groups defined by the size of their contributions. I.e., the more popular a book is, and the more its relative popularity fluctuates between months, the higher its contribution (see Methods for details). Groups are non-overlapping and defined as: i) the top 100 most contributing books, ii) the top 101 to 1000 most contributing books, iii) top 1001 to 10.000, iv) top 10.001 to 50.000, and v) books beyond top 50.000. Using these groups, we explore how many books we need to account for a given proportion of the drift over time, and how consistent the contributions of individual books are.

\begin{figure}[!htbp]
\centering
\includegraphics[width=0.9\textwidth]{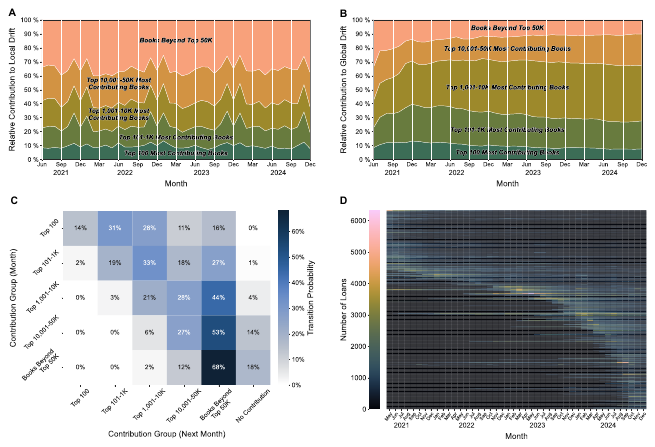}
\caption{\textbf{Contributions of individual books to drift dynamics.}
\textbf{A-B,} Distributions of relative contributions of individual books to local drift (\textbf{A}) and global drift (\textbf{B}) over time. \textbf{C,} Probability that a book with a given level of contribution to the local drift in a month will have a certain level of contribution to the local drift in the next month. Probabilities are averaged across all pairs of months. \textbf{D,} Monthly popularity for the 1,000 books with the highest contributions to the global drift in October 2024. Books are vertically ordered by the month where their popularity peaks.}\label{fig_3}
\end{figure}

We start by examining how much each group of books contributes to the drift in collective attention over time (Fig.~\ref{fig_3}).
Surprisingly, we find that the 100 books with the most significant shifts in their popularity consistently account for only around 10\% of both local and global drift (Fig.~\ref{fig_3}A-B). This means that, even if it were possible to identify these books and predict changes in their popularity, 90\% of the total shift in collective attention would remain unexplained without also accounting for less influential books. 
For local drift, the top 10,000 books that contribute most to drift account for less than 50\% of the total change in collective attention.
When it comes to seasonal variation in local drift (Fig.~\ref{fig_3}A), we observe a slight increases over time in the relative contributions of the top 101-1,000 highest contributing books, whereas the top 100 most contributing books do not increase their contributions.
For global drift, we find that the relative contributions of the top 1,000 books decline over time (Fig.~\ref{fig_3}B). In other words, the 1,000 books with the largest changes in popularity are less important for explaining collective attention drift, the further we move from the reference month. Instead, shifts in the popularity of especially the top 1,001-10K most contributing books account for a larger proportion of global drift over time.
Nonetheless, even when accounting for the top 10.000 most influential books, we explain less than 70\% of global drift.
Taken together, these results show that books in the long tail of the popularity distribution, in a collective fashion, are impactful in driving changes in collective attention.

Although each group's contribution to local drift is relatively stable across months, it is unclear to what extent books transition between the different contribution groups (top 100, top 101-1,000, etc.) over time. To explore this, we measure the likelihood that a book belonging to a particular contribution group in a given month would remain in the same group or transition to one of the other groups in the following month. We average these transition probabilities over all pairs of consecutive months, and find that only 14\% of the 100 books that contribute most to local drift in one month are also part of the 100 highest contributors in the next month (Fig.~\ref{fig_3}C). This means that a very large shift in a book's popularity are typically followed by a smaller shift. 
Taken together, individual books show substantial variation in their contributions to drift of collective attention between consecutive months.

To understand which books drive long-term shifts in collective attention, we examine the 1,000 books contributing most to global drift in October 2024 – the last month in the data unaffected by the observed seasonal patterns – and order them by when they reach peak popularity (Fig.~\ref{fig_3}D). Two patterns emerge. First, rather than distributing evenly across our observation period, the peak popularity of these books cluster at the boundaries of our observation window: many peak shortly after the reference month (May 2021), while others peak approaching October 2024. This ordering reveals that global drift arises from two distinct sources: books whose popularity has faded since early in our study period and those gaining recent prominence. Second, the popularity of these high-contributing books is remarkably ephemeral. These two patterns – clustering at the edges of the observation period and volatile popularity trajectories – distinguish books driving the global drift from the most popular books and books with highest peaks in popularity (Supplementary Figs.~S26-27), revealing that standard measures of book popularity do not fully capture which titles drive long-term collective attention shifts. In Supplementary Fig.~S28, we show that the seasonal drift, identified above, are largely driven by specific Christmas-themed children’s titles which regain popularity each holiday season but remain relatively `unpopular' during the rest of the year.

In summary, we find individual books to have limited power for explaining drift in collective attention. As such, even if we could perfectly predict the popularity trajectories of the books with the largest changes in popularity, they would only account for a small percentage of the total drift in collective attention. This underscores the importance of considering the full `cultural system' and not just individual product trajectories when analyzing patterns of consumer attention.

\subsection*{Generalizability and predictability of drift patterns}
So far, we have shown that the collective attention of library loaners is constantly changing. This has a direct impact on the predictability of the system. Further, it implies that the performance of any machine learning model fitted on the cultural consumption at a given point in time, e.g., to recommend books, will deteriorate over time if not continually kept-up-to-date and retrained. However, since drift in collective attention exhibits stable patterns, it may be possible to predict attention drift and get an estimate on how fast model performance will decay and how often models should be retrained.

\begin{figure*}[!htbp]
\centering
\includegraphics[width=\textwidth]{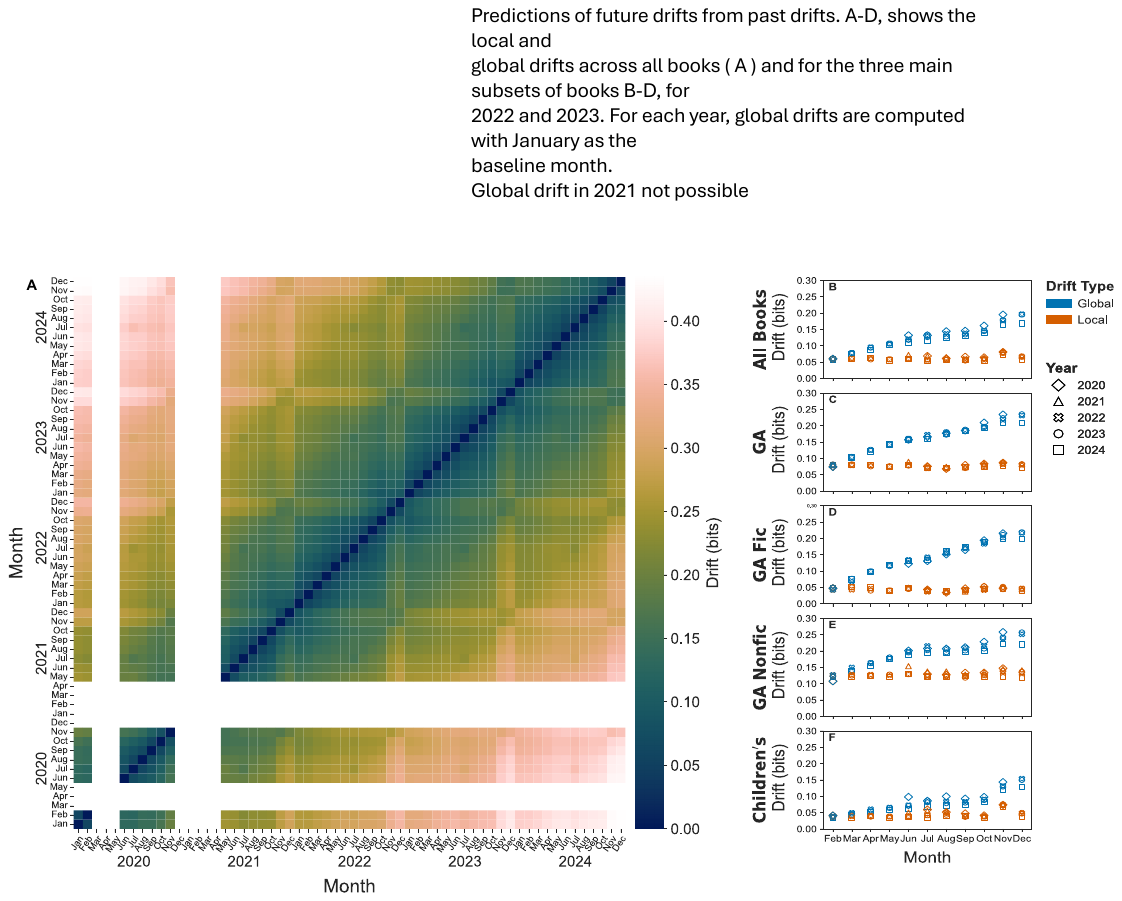}
\caption{
    \textbf{Predictability of collective attention drift across years.}
    \textbf{A} shows all pairwise drifts between months from 2020 to 2024 (excluding COVID-affected months), illustrating the increase in drift with temporal distance and the consistency of seasonal patterns.
    \textbf{B} shows the local and global drift for all books, and \textbf{C–F} show the corresponding drift for the major book subcategories (GA stands for General Audience). 
    For each year, global drift is computed using January as the baseline month. Since January 2021 was affected by a partial COVID-19 lockdown, global drift is not computed in 2021.
}
\label{fig_4}
\end{figure*}

To examine whether drift patterns are stable over time, we quantify drift between all pairs of months over the five-year period from 2020 to 2024, excluding months affected by COVID-19 lockdowns. A visual inspection of drifts in Fig.~\ref{fig_4}A shows that the patterns we previously identified, for a subset of months, are remarkably consistent across all months; drift in collective attention generally increases with distance in time between any two months, and identical seasonal patterns appear across all years. In the SI, we show that the drift patterns are equally consistent for book subcategories and that the pattern of increasing drift over time also holds during COVID-19 lockdown months, although drift in these months are generally higher (Supplementary Figs.~S29-33). In Fig.~\ref{fig_4}B–F, we further show that both local and global drift remain highly stable from year to year, for all books and across the different book categories included in our analysis. This stability suggests that future changes in collective attention can be approximated from past changes.

We now consider whether drifts in future years can be predicted using observations from earlier years. For this purpose, we use a minimal model. For any pair of months in a given year, we use the drift observed for the same period in the previous year as the prediction. For example, if attention drifted by 0.2 bits between two months in 2022, we predict that the attention between the same two months in 2023 will also drift by 0.2 bits. Despite its simplicity, this model performs remarkably well. Across all loaner subgroups, the mean absolute percentage error (MAPE) for predicting one year’s drift from the preceding year remains below 10\% with slightly lower errors for general audience books than children's books (Supplementary Figs.~S34-38). Prediction accuracy degrades when using older reference years for making predictions, suggesting subtle long-term changes in drift patterns.

Together, these results show that, although we cannot predict which specific items will gain or lose attention, we can reliably predict the magnitude of drift in collective attention. This has important implications for policy makers, regulators, and practitioners dealing with how often algorithmic systems that recommend cultural items to users should be updated and evaluated.

\section*{Discussion}
For decades, studies of cultural markets have emphasized the difficulty of predicting the ever-changing attention of cultural consumers~\cite{caves_creative_2000, vany_hollywood_2003, salganik_experimental_2006, salganik_leading_2008, watts_common_2014}. We show that changes in the collective-level attention of Danish book loaners are consistently high and cannot be attributed to only a subset of titles. However, our study shows that the collective redistribution of attention across books in Danish libraries has identifiable global patterns that can be predicted with a high degree of accuracy, even for specific groups of products and audiences. While we examine only broad product and audience groups, our findings suggest that predictable collective attention dynamics could exist for more fine-grained groupings, such as sub-genres. 

Our findings raise important questions about the mechanisms underlying collective cultural change, specifically processes of individual choice and social influence. Relationships between individual choice, social influence, and cultural change have previously mainly been investigated in simulation studies~\cite{axelrod_dissemination_1997, muthukrishna_are_2020, jung_cultural_2021, acerbi_weak_2025} and experimental work~\cite{salganik_experimental_2006, salganik_leading_2008, epstein_social_2021, hentenryck_aligning_2016, krumme_quantifying_2012, abeliuk_taming_2017}. 
Detailed registry data on individual loaners' choices, such as the one we used in this study, can open new opportunities to study relationships between behavior at the individual-, group-, and macro-levels, especially when combined with administrative data on the social opportunity structures of each individual loaner~\cite{cremers_unveiling_2025}.

Inferring mechanisms from purely observational data, however, requires caution. Although simple mechanistic models – such as those based on collective memory~\cite{candia_universal_2018}, neutral drift~\cite{bentley_regular_2007}, and information cascades~\cite{bikhchandani_theory_1992} – have been shown to reproduce patterns of cultural change in empirical data, such patterns are often compatible with multiple alternative explanations~\cite{gelastopoulos_reinforcement_2025, leroi_neutral_2020, johnson_curvefitting_2025}. We have avoided imposing any particular model of individual-level selection and instead focused on the patterns in the collective attention that emerge from loaners' aggregated choices. We have shown that these patterns are both interpretable and predictable without knowing the underlying mechanisms. At the same time, they place constraints on which mechanisms could plausibly generate them. For example, the recurrence of holiday-themed children's books indicates that models which do not account for fluctuating selections cannot reproduce the observed dynamics. That such fluctuations are observed for children's books but not general audience books further suggests heterogeneity in the mechanisms driving selection across different book types. Heterogeneity of selection mechanisms is incompatible with models based on pure reinforcement like neutral models, which have been used extensively for to explain cultural change~\cite{bentley_random_2004, bentley_regular_2007}.

The extent to which our results can be generalized to other cultural markets is an empirical question, however, we expect similar forms of collective attention dynamics to apply widely. For instance, other cultural markets display variations of the drift components we observe for library loans: a stable turnover in consumer preferences and season-specific item consumption~\cite{bentley_regular_2007, park_global_2019}. Nonetheless, the importance of individual products for drift dynamics could differ significantly between markets. In markets with fewer products and more unequal distributions of popularity, a small number of products may account for a larger proportion of drift. In such markets, shocks by individual products could disrupt otherwise stable drift patterns, causing more unpredictable drift dynamics than what we observe for library loans. We hope that future studies will explore how characteristics of drifts in collective attention compare across various cultural markets.

Our dataset contains comprehensive records of all Danish public library loans; however, our findings are not guaranteed to generalize across all book consumption. Book consumption is most likely different when there is an associated economic cost. For example, borrowing books at the library is free which may cause users to explore titles more widely compared to when they have to buy books in a bookstore. Similarly, libraries feature a wide selection of both new and older titles, whereas bookstores prioritize newer releases and popular titles. These factors could likely influence the drift of collective attention.

Recommender systems are another potential source of variation in consumer attention dynamics between markets. Recommender systems are pervasive in many current cultural markets, and their effects on drift in collective consumer attention should be studied more closely. However, because these systems are so deeply embedded in most real cultural markets, isolating their effects is a challenge~\cite{wagner_measuring_2021}. Even public libraries rely heavily on centralized recommendation systems, both on their websites for book search and reservations, and via librarians’ curation of books in each local library.

Our findings have major implications for adapting to changing consumer attention. With reliable expectations of drifts in consumer attention, individuals and organizations interested in predicting and modeling cultural consumption can make more informed assessments of when machine learning models require updating. As we show, drifts in collective attention can be faster for certain product types and consumer groups (Fig.~\ref{fig_2}). Model updating should take these group-specific drift differences into account. Failing to do so could result in problematic biases towards groups with faster drift in collective attention. For example, in our data, drift is higher for women and for 65+ year-old demographics, meaning models need to be updated more frequently to accurately reflect their tastes. The consequences of drift in collective attention will depend on the specific model. In particular, many recommender systems capture associations between users and items~\cite{su_survey_2009}. In these models, reductions in performance can result not only from drift in item popularity but also from drifts in the user base and drifts in the associations between users and items. Our study focuses solely on item popularity, leaving open important questions about the predictability of drift in other data used to model cultural consumption.

We envision the methods developed in this study to be used to study changes in collective behavior in other domains outside of cultural markets. Like cultural consumption, behaviors that are socially driven are generally difficult to predict, including information cascades, social influence, and social interactions~\cite{cheng_can_2014, hofman_prediction_2017, bikhchandani_theory_1992, dedomenico_interdependence_2013}. For such behaviors, it may be possible to identify and predict stable macro-level patterns of changing behavior that can be used to contextualize and adapt to uncertain predictions of individual phenomena, like the popularity of specific topics or items.

\section*{Methods}
\subsection*{Ethics and anonymity}
Since this study was based on a secondary analysis of administrative registers, no ethical approval was required for this study. All data processing and analysis was conducted in-house at \emph{Statistics Denmark}, the national statistical authority of Denmark, in compliance with their data protection protocols and the requirements of the Danish Data Protection Agency. This means that all individual-level data, both the library takeout dataset and the sociodemographic loaner information, were pseudonymized and kept strictly confidential. 

\subsection*{Data Access}
Due to the sensitive nature of the data, it is not possible to openly share the data, but it is possible to apply for access to the data through Statistics Denmark.

\subsection*{The Danish Nation-Scale Library Takeout Dataset}
The library dataset used in this study is based on administrative registers of library takeouts from Danish public libraries, collected by \emph{Statistics Denmark}. The dataset is continuously updated, and the version used in this study spans from January 1, 2020, to December 31, 2024, and includes all loans from public libraries, including \emph{Ereolen} (the Danish national platform for digital books), but excluding university and school libraries.

\subsection*{Sociodemographic Loaner Information} 
All variables except sex are dynamic in our dataset. This means that loaners can belong to different subgroups at different points in time. Below, we describe how we derived the different loaner categories.
\textit{Age} is computed directly from each loaner's birthdate at the time of each loan.
\textit{Residence Area} is based on standard definitions provided by \emph{Statistics Denmark}, which divides municipalities into five types (\url{https://www.dst.dk/en/Statistik/dokumentation/nomenklaturer/kommunegrupper}), which we collapse into two main types: 1) \emph{Large city municipality}, which is defined by a municipality with above 100,000 citizens or has a high job availability (see \url{https://www.dst.dk/en/Statistik/dokumentation/nomenklaturer/kommunegrupper} for specifics), and 2) \emph{Town or rural municiplaity}, which covers all other municipalities.
\textit{Level of Education} is recorded as the loaner's maximum attained education level at the time of each loan. The three categories are based on the DISCED-15 classification (see \url{https://www.dst.dk/en/Statistik/dokumentation/nomenklaturer/disced15-udd}), which groups education into 15 main categories.

\subsection*{Library Data Preprocessing} 
\subsubsection*{Item Filtering}
Around 12 million loans in the dataset refer to non-books (see Supplementary Fig.~S2 for the distribution of loans across material types), which were removed from analysis.

\subsubsection*{Book Identification}
Each item is associated with a unique identification number, known as a \emph{faust number}. Because each edition of a book gets a separate \emph{faust number}, one title can have several associated \emph{faust numbers}. Since we are interested in the evolution in the popularity of titles, regardless of the medium (physical book, audiobook, etc.), we treat different media variants of the same title as a single entity. To derive a unique identifier independent of the edition and medium, the following heuristic algorithm was applied:
\begin{enumerate} \scriptsize
    \item Remove special characters from book titles, and make all characters in the titles lowercase.
    \item Sort all book titles and authors lexicographically.
    \item For each book, compare it with the succeeding ten books in the sorted list, and find all pairs of books that satisfy the following criteria:
    \begin{enumerate}
        \item The edit distance between their titles and authors is at most 1.
        \item Both book titles contain the same digit of maximum length 2, indicating a particular version of the book. Book titles with no digits and books with a digit of 1 in the title are treated as the same (e.g., \emph{Ternet Ninja} and \emph{Ternet Ninja 1}).
    \end{enumerate}
    \item For all pairs that follow these criteria, group together any pair where one of the books overlaps.
\end{enumerate}
This process reduced the unique number of items from approximately 1,060,000 to around 760,000.

\subsubsection*{Book Category}
Book categories (general audience fiction, general audience nonfiction, children's) were provided by the \emph{Danish Bibliographic Centre}.

\subsection*{Measuring Changes in Collective Consumer Attention} 

\subsubsection*{Jensen-Shannon Divergence and Book-Level Contributions}
In each month, we compute the popularity (number of loans) of each book $i$ relative to the total amount of loans in that month. The drift in collective attention between those months is the JSD between the distributions of popularity in two months $P$ and $Q$, which can be expressed as a sum of individual book contributions:
\begin{align*}
    \text{JSD}(P, Q) &= \sum_i \Bigl( \overbrace{{0.5p_i \log \frac{2p_i}{p_i+q_i}} + {0.5q_i \log \frac{2q_i}{p_i+q_i}}}^{\text{Partial JSD}_i} \Bigr)
\end{align*}
where $p_i$ is the relative popularity of book $i$ in the first month, $q_i$ is the relative popularity of book $i$ in the second month. The partial JSD represents the contribution of each book to the overall JSD. If there is no change to a book's relative popularity, it does not contribute to the JSD. For any measure of drift, the relative contribution of single book to that drift is simply calculated as the partial JSD of that book divided by the total JSD.
\\

\subsubsection*{Estimation of Jensen-Shanon Divergence}
Empirical estimation of information-theoretic quantities from samples is known to suffer from bias, particularly when the number of categories (here: unique books) is larger than the total number of sampled counts (here: loans) \cite{paninski_estimation_2003}. To maintain a consistent relation between changes in overall collective attention and changes in collective attention for individual books, we use the maximum likelihood estimator when computing JSDs for all loaners.

To ensure robust comparisons between sociodemographic subgroups, we compute JSDs only for the top 10,000 books in each group. We use a bootstrap estimator, which has been shown to outperform other estimators in cases where the number of counts is higher than the number of categories \cite{dedeo_bootstrap_2013}. Additionally, this estimator provides uncertainty estimates for each computed JSD. 500 bootstrap permutations were used to compute each bootstrap estimate.

\backmatter

\bmhead{Acknowledgements}
We are grateful for the help with data collection by the Data Science Lab at Statistics Denmark and Ea Hoppe Blaabæk.

\section*{Competing interests}
The authors declare no competing interests.

\section*{Author contribution}
Both authors designed the study, WL performed the analysis, both authors analyzed the data, and wrote the manuscript.

\bibliography{sn-bibliography}

\end{document}

\title[Article Title]{Predictable Drifts in Collective Cultural Consumption: Evidence from Nation-Level Library Takeout Data}
\subtitle{\emph{Supplementary Information}}

\author*{\fnm{Anders} \sur{Weile}} %\email{awel@itu.dk}
\author*{\fnm{Vedran} \sur{Sekara}} %\email{vsek@itu.dk}

\maketitle

\hypersetup{
    linkcolor=black % Set link color to black
}
\hypersetup{
    linkcolor=blue % Reset link color to blue
}

\newcommand{\SIref}[1]{SI~\nameref{#1}}

\begingroup
\renewcommand{\thefigure}{S\arabic{figure}}
\setcounter{figure}{0}

\section*{Summary of Supplementary Information}
The Supplementary Information begins with four short sections on respectively alternative drift measures (Supplementary Section 1.1), the results from a null model of stochastic drift (Supplementary Section 1.2), the identification of Christmas books used in the analysis (Supplementary Section 1.3), and some examples to illustrate the connection between loans, local drift, and global drift (Supplementary Section 1.4),  and . In addition, this document contains supplementary figures to accompany the main results of the article.

\subsection*{Supplementary Section 1.1 Generalized Jensen-Shannon Divergence for measuring the robustness of drift patterns across the entire spectrum of popularity}
The Generalized $\alpha$-Jensen-Shannon Divergence ($\text{JSD}_\alpha$) is a single-parameter generalization of the JSD, where $\alpha$ controls the emphasis on changes to more popular books ($\alpha > 1$) and less popular books ($0 \leq \alpha < 1$). $\text{JSD}_{\alpha=1}$ is equal to the standard JSD, where popular and unpopular books are weighted equally. $\text{JSD}_{\alpha=0}$ simply measures the fraction of books that are different in two time periods, regardless of how many times those books have been loaned. We use the ($\text{JSD}_\alpha$) to measure the robustness of the patterns we observe in the main results across the entire spectrum of book popularity. In the main text, we operationalize the JSD as a sum of partial contributions by individual books, but the JSD can also be expressed in terms of Shannon Entropy $H(\cdot )$:
\begin{align*}
    \text{JSD}(P, Q) = H(M) - \frac{H(P) + H(Q)}{2}
\end{align*}
where $P$ and $Q$ (in this context) are distributions of relative popularity over all books in the dataset, and $M$ is a mixture distribution of $P$ and $Q$: $M = (P + Q) / 2$. 

Similarly, $\text{JSD}_\alpha$ can be defined in terms of Tsallis entropy of order $\alpha$, $H_{\alpha}(P)=\frac{1}{1-\alpha}(\sum_i p_i^{\alpha}-1)$ \cite{havrda_quantification_1967}:
\begin{align*}
    \text{JSD}_\alpha(P, Q) &= H_\alpha(M) - \frac{H_\alpha(P)+H_\alpha(Q)}{2}
\end{align*}
To make the $\text{JSD}_\alpha$ comparable to the JSD, we use the normalized version, defined in \cite{gerlach_similarity_2016}:
\begin{align*}
    \overset{\sim}{\text{JSD}}_\alpha(P, Q) &= \frac{\text{JSD}_\alpha(P, Q)}{\text{JSD}_\alpha^{\text{max}}(P, Q)}
\end{align*}
where
\begin{align*}
    \text{JSD}_\alpha^{\text{max}}(P, Q) = \frac{2^{1-\alpha}-1}{2} \left(H_\alpha(P) + H_\alpha(Q) + \frac{2}{1-\alpha}  \right)
\end{align*}
$\text{JSD}$ and $\overset{\sim}{\text{JSD}}_\alpha$ are both symmetric and bounded between 0 (if and only if $P=Q$) and 1 (no overlap in borrowed titles). $\sqrt{\text{JSD}}$ is a true metric \cite{endres_new_2003}, while $\sqrt{\text{JSD}_\alpha}$ is a metric for $\alpha \in (0, 2]$ \cite{briet_properties_2009}, and $\sqrt{\overset{\sim}{\text{JSD}}_\alpha}$ is a metric for $\alpha \in (0, 2]$ when the same rank-frequency distribution underly all $Ps$ and $Qs$ \cite{gerlach_similarity_2016}. As an alternative measure of overlap that does not consider book popularity to $\sqrt{\overset{\sim}{\text{JSD}}_{\alpha=0}}$, we use the Jaccard Distance (JD), which is a true metric \cite{kosub_note_2019}:
\begin{align*}
    \text{JD}(P, Q) = 1 - \frac{|\text{Supp}(P) \cap \text{Supp}(Q)|}{|\text{Supp}(P) \cup \text{Supp}(Q)|}
\end{align*}

\subsection*{Supplementary Section 1.2 Null Model}
For our main results, we use the number of monthly loans of a book as a direct measure of the probability that the book is loaned in that month. This choice implies that part of the observed drift may arise from random variation in loan counts rather than from meaningful shifts in collective attention. To assess the extent to which such random fluctuations could generate drift patterns similar to the empirical ones, we construct a null model following~\cite{gerlach_similarity_2016}. This model assumes that the loans in two months are generated from the same underlying probability distribution, meaning that all changes between the months are purely stochastic. Concretely, for each pair of months, we construct an average distribution of loans $p' = 0.5 p_1 + 0.5 p_2$, where $p_1$ and $p_2$ are the empirical probability distributions over book loans in the two months. We then sample $n_1$ and $n_2$ loans, corresponding to the total number of loans in the two months, compute the associated probability distributions of these samples, and measure the Jensen–Shannon divergence (JSD) between them. For each pair of months, we compute the JSD over 500 independent samples. In SI Fig.~\ref{null_model}, we show the local and global drifts in the null model for the five book categories. In the null model, the local drifts are noticeably lower than in the real data, and the global drift remains constant over time. Moreover, no November–December seasonal effects appear. These results demonstrate that the observed increase in drift over time and the seasonal patterns cannot be explained by random fluctuations.

\begin{figure*}[h!] %[H]
\centering
\includegraphics[width=0.9\textwidth]{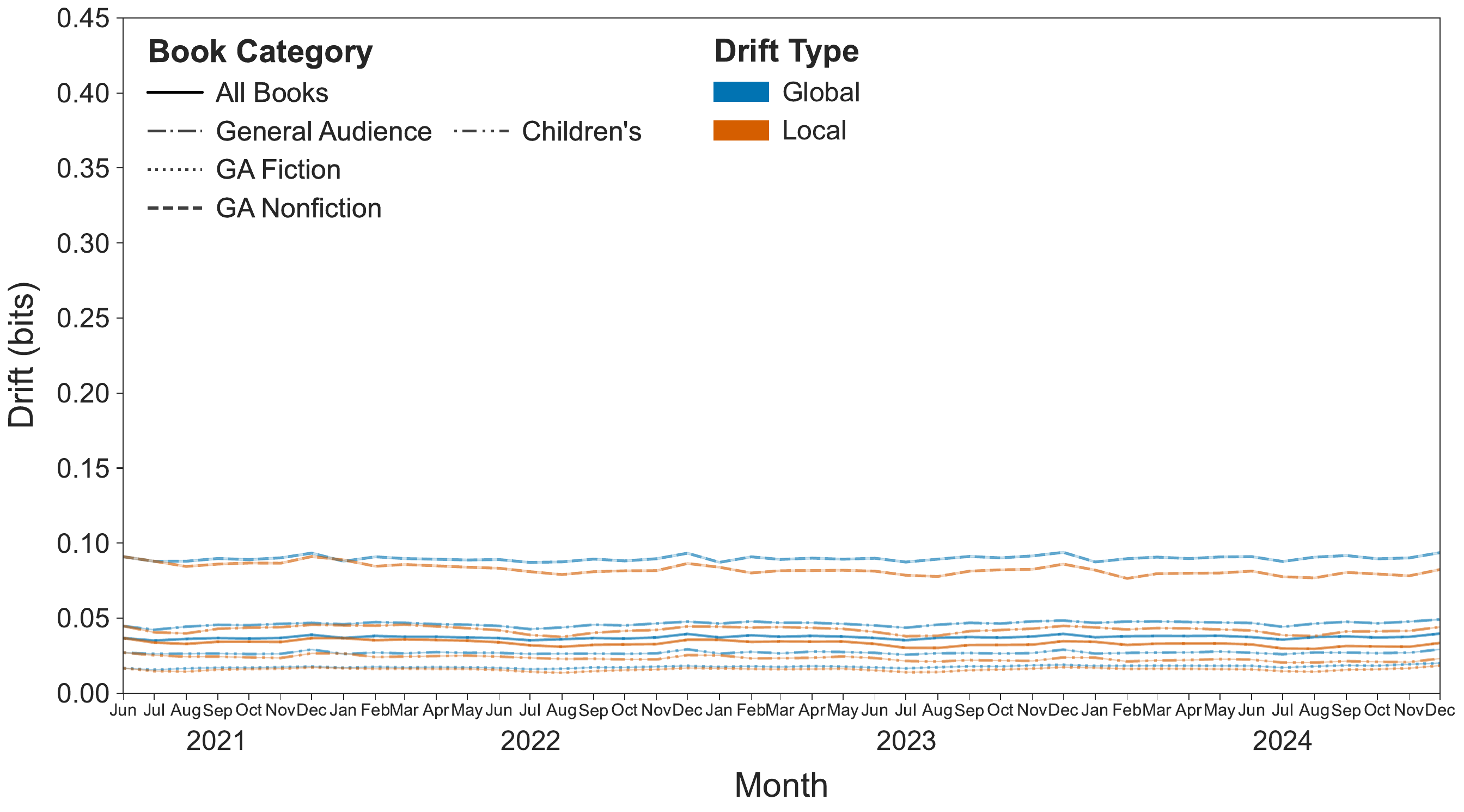}
\caption{
    Local and global drifts in a null model that assumes that loans in two months are samples from the same underlying distribution. Shaded areas around lines (barely visible) show $2\sigma$ bootstrap errors.
}
\label{null_model}
\end{figure*}

\subsection*{Supplementary Section 1.3 Partial JSD examples}
We illustrate the dynamics at play for three concrete books, illustrating how the evolving attention from consumers towards these books relates to their contributions to local and global drifts over time (SI Fig.~\ref{individual_contributions_examples}A-C):
\begin{itemize}
    \item 'Annas Sang' ('\emph{Anna's Song}') was the most loaned book in the entire dataset, even though it was not published until May 22, 2022. After its release, the book quickly rose in popularity, reflected in a high contribution to the local drift between May and June 2022. During the following year, it became increasingly popular among library loaners until reaching peak popularity in June 2022, after which its popularity started to decline. Because this book was released after the baseline month for computing the global drift (May 2021), the global drift contribution matched its relative popularity in a given month.
    \item Another popular book in this period, 'Meter i sekundet' ('\emph{Meters per Second}') was published in 2020 and later made into a movie, which was released on February 2, 2023. Since it was already popular by May 2021, 'Meter i sekundet' did not contribute as much to drifts in attention as 'Annas sang', even though they were comparably popular. The release of the movie had a clear impact on the attention towards the book, which is seen as a sudden increase in the otherwise steadily declining popularity. As a result, the book's local drift contribution was high in January and February 2023. Since its popularity got closer to its baseline popularity in May 2021, its global drift contribution was relatively low,
    \item For the highly popular Christmas book "Sallys far kringler julen" ('\emph{Sally's Dad Fixes Christmas}'), popularity was highly seasonally dependent. Mirroring the seasonal patterns of local and global drift in the total consumer attention, this book's contributions to local drifts were high in November-January, and contributions to global drifts were high in November-December. 
\end{itemize}
As these examples highlight, books can have different popularity trajectories and consequently contribute to drifts in diverse ways over time.

\begin{figure}[H] %[H]
\centering
\includegraphics[width=1.0\textwidth]{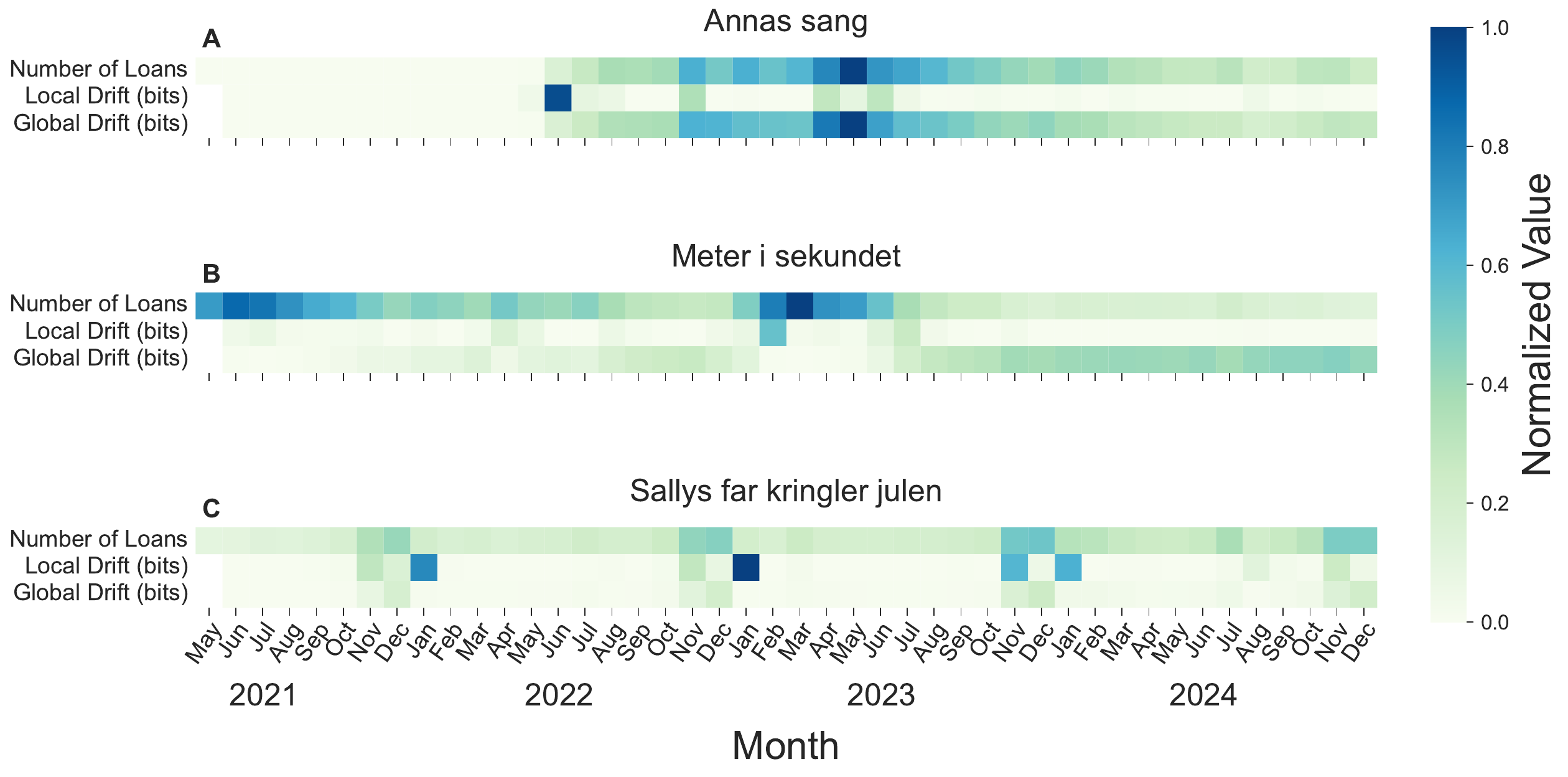}
\caption{
Loan and drift contribution trajectories of three select books. For comparison purposes, each of the three variables (number of loans, local drift, global drift) is divided by the maximum recorded value for that variable across the three books.
}
\label{individual_contributions_examples}
\end{figure}

\subsection*{Supplementary Section 1.4 Impact of Christmas books on Drift}
We identify Christmas books using two criteria: (1) books listed as Christmas-themed on bibliotek.dk, the Danish public library portal, and (2) books with titles containing the Danish word \emph{jul} (Christmas) or the prefix \emph{jule-}, e.g., \emph{juleeventyr} (Christmas adventure) and \emph{julehistorie} (Christmas Story). Together, these criteria yield 3,225 unique Christmas books across all categories out of the about 760,000 total unique books. As shown in SI Fig.~\ref{christmas_unique_books}, the identified Christmas books are mostly children's books, constituting 1.3\% of all unique titles in the children's and young adult category, compared to 0.6\% of general audience fiction and 0.2\% of general audience nonfiction. They also account for a small shares of the total loans in each category (SI Fig.~\ref{christmas_num_loans}).

Christmas books and non-Christmas books show very different patterns in how their popularity evolve. The most popular non-Christmas books (SI Fig.~\ref{most_popular_non_christmas_books}) usually have single peaks in their popularity but otherwise maintain relatively stable increases or decreases in popularity, except each December where their popularity suddenly drops. By contrast, the most popular Christmas books (SI Fig.~\ref{most_popular_christmas_books}) show pronounced spikes in November-December with near-zero loans in the remaining months. When drift is computed separately for each subset (SI Fig.~\ref{drift_christmas_subset}), Christmas books drive the pronounced seasonal peaks in both local and global drift, while the peaks almost disappear for the non-Christmas subset.

\begin{figure}[H]
\centering
\includegraphics[width=0.9\textwidth]{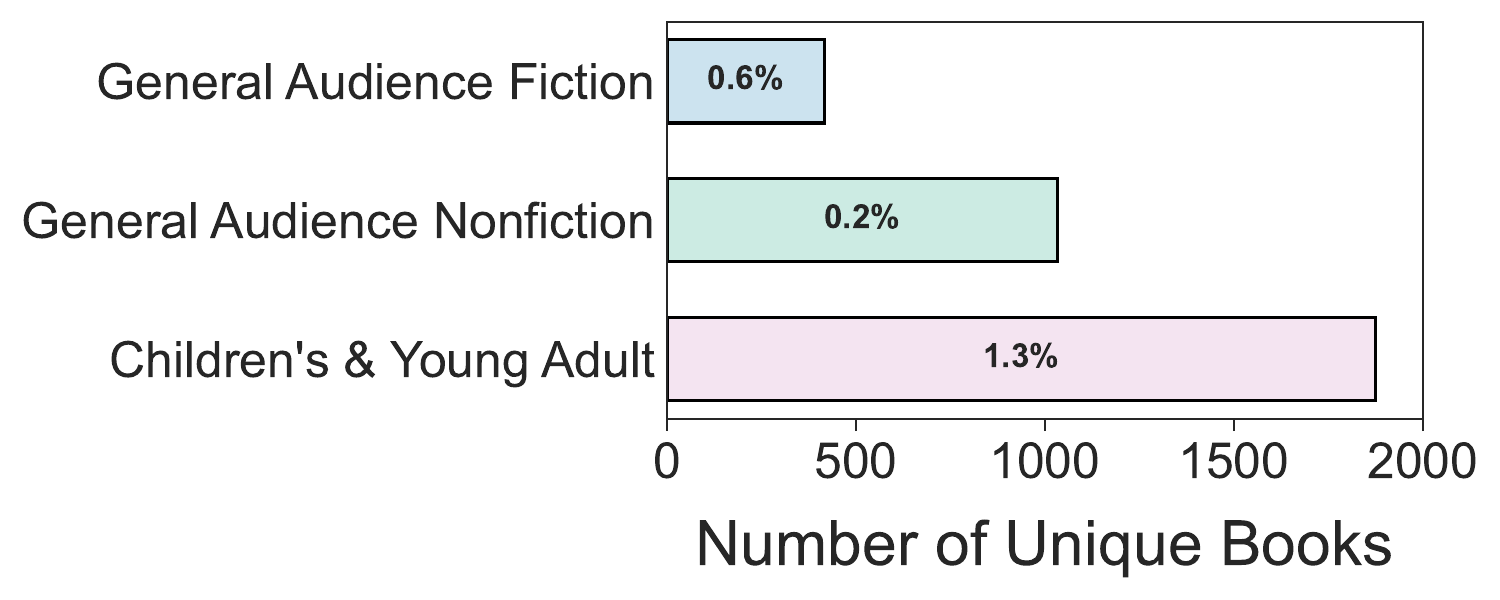}
\caption{
    Number of unique Christmas book titles by book category. Percentages indicate the share of all unique titles within each category that are classified as Christmas books.
}
\label{christmas_unique_books}
\end{figure}

\begin{figure}[H]
\centering
\includegraphics[width=0.9\textwidth]{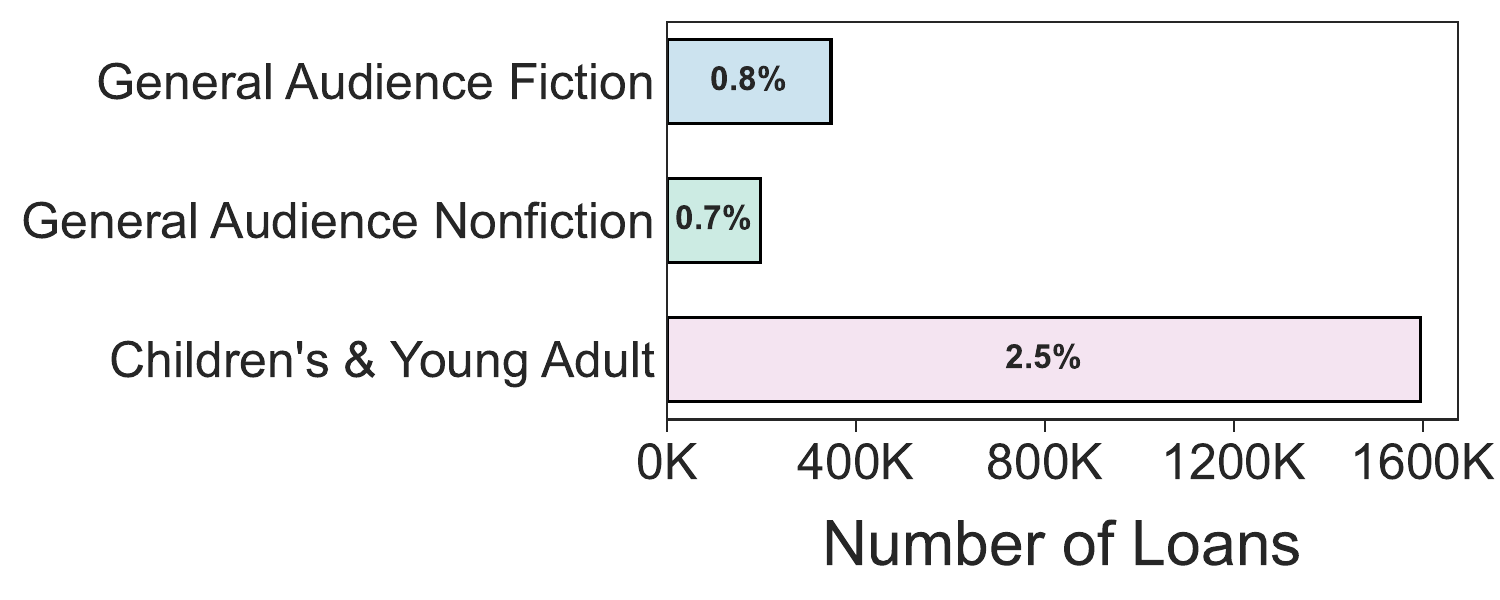}
\caption{
    Total number of loans attributed to Christmas books by book category. Percentages indicate the share of all loans within each category that are accounted for by Christmas books.
}
\label{christmas_num_loans}
\end{figure}

\begin{figure*}[h!]
\centering
\includegraphics[width=0.8\textwidth]{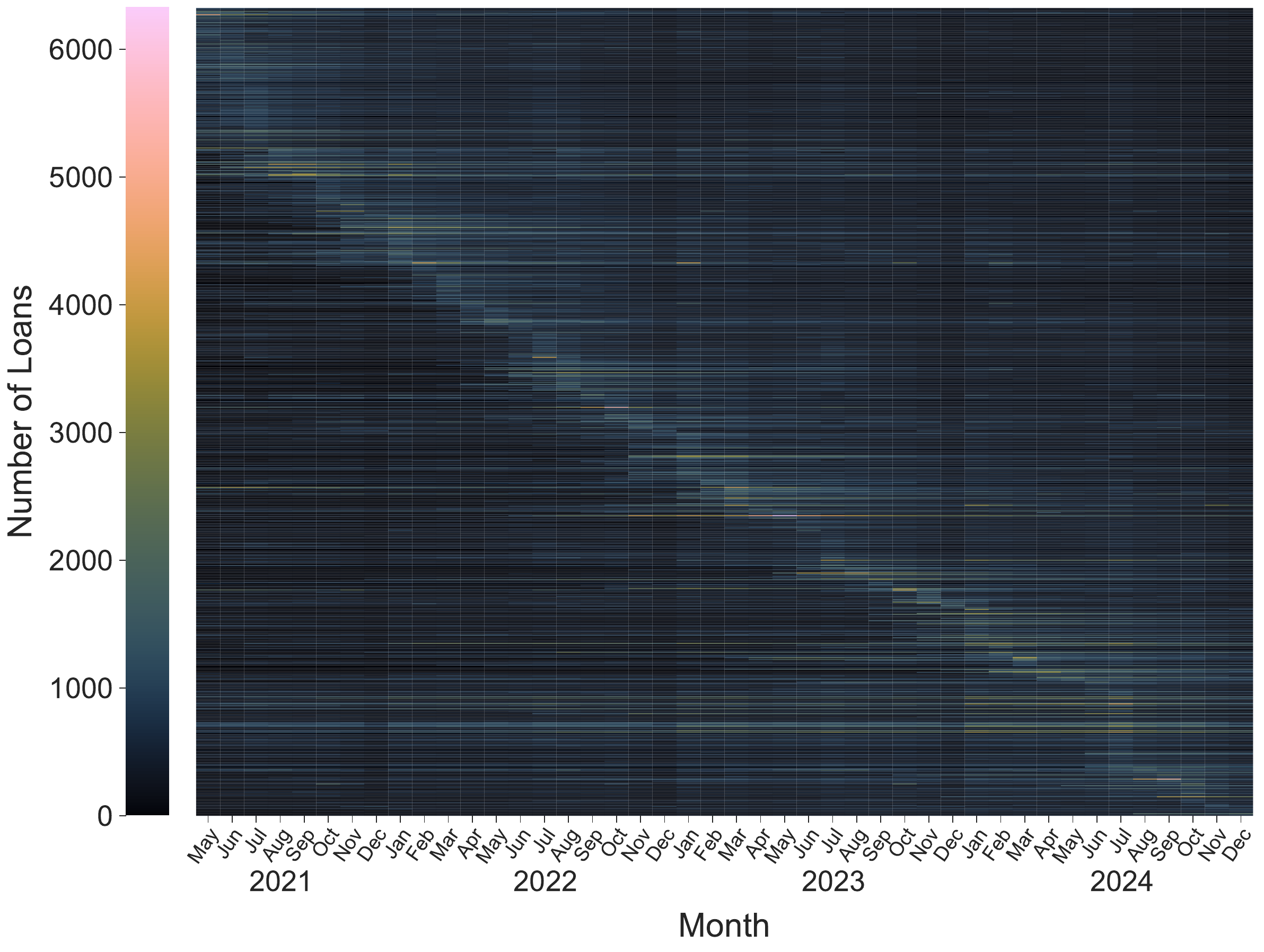}
\caption{
    Monthly popularity for the most popular non-Christmas books between May 2021 and December 2024, ordered by the month where their popularity peaks.
}
\label{most_popular_non_christmas_books}
\end{figure*}

\begin{figure*}[h!]
\centering
\includegraphics[width=0.8\textwidth]{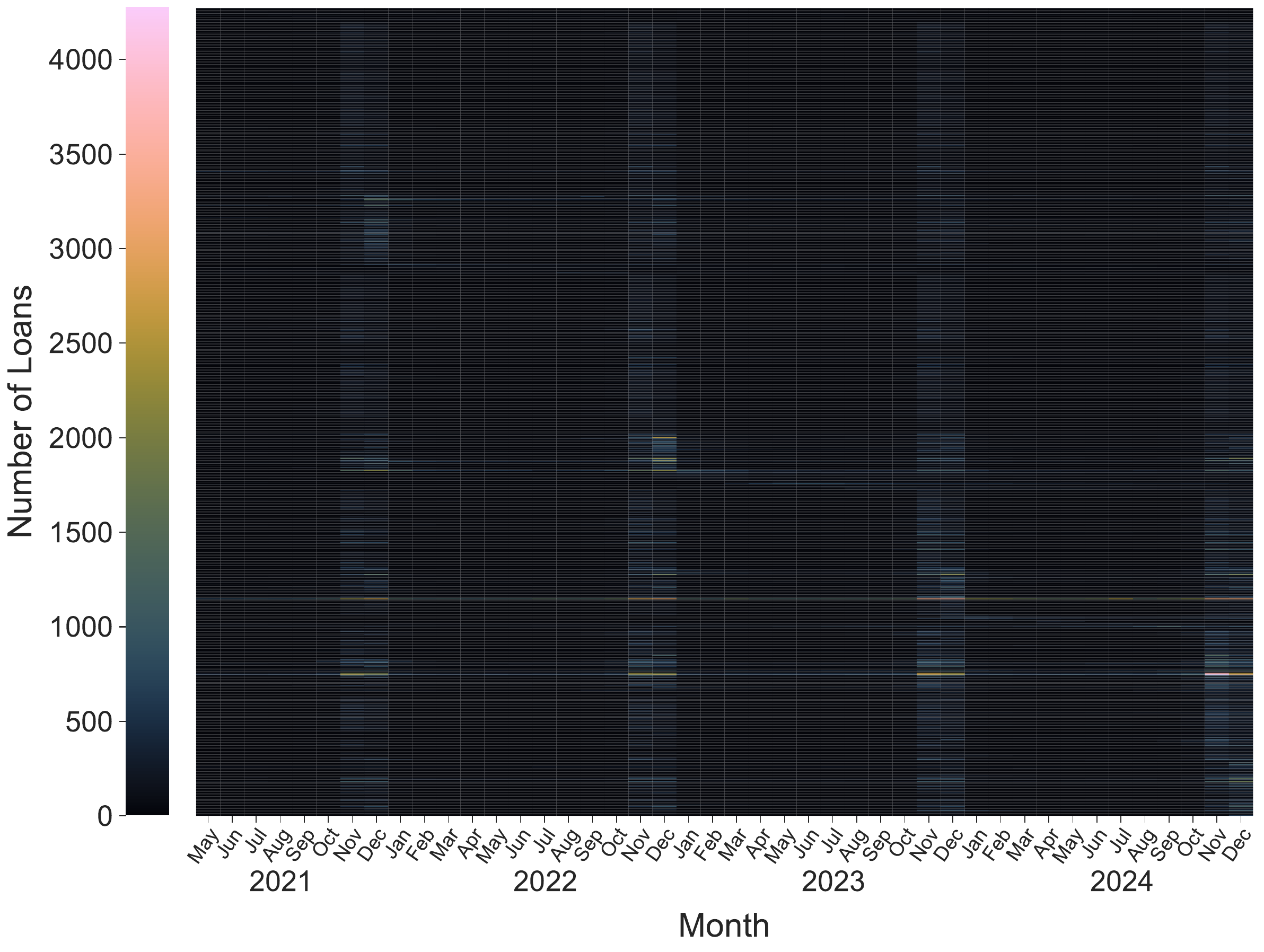}
\caption{
    Monthly number of loans for the most popular Christmas books between May 2021 and December 2024, , ordered by the month where their popularity peaks. Loans are strongly concentrated in November-December of each year, with near-zero activity outside the Christmas season.
}
\label{most_popular_christmas_books}
\end{figure*}

\begin{figure*}[h!]
\centering
\includegraphics[width=0.9\textwidth]{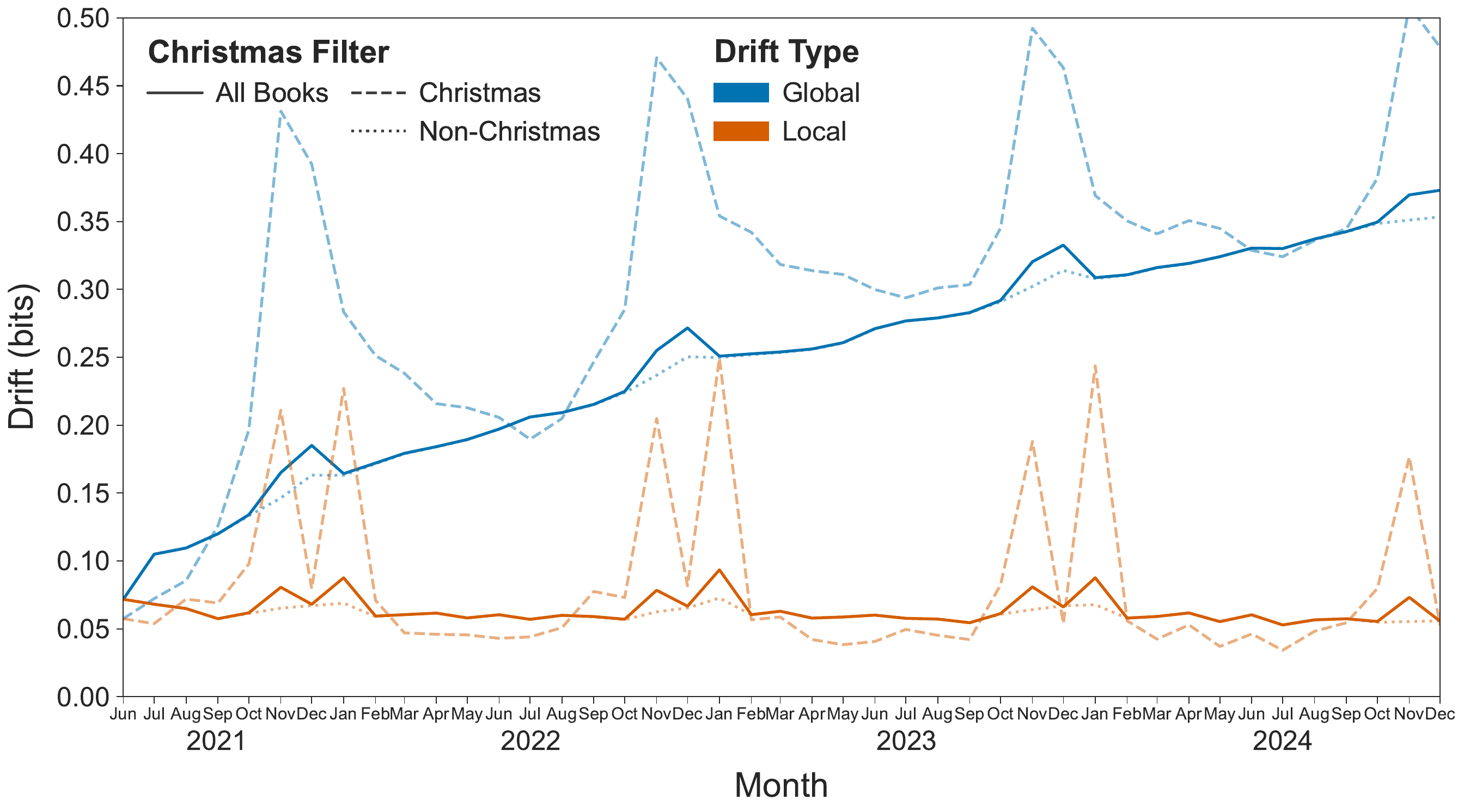}
\caption{
    Local and global drift in collective attention for all books (solid lines), Christmas books (dashed lines), and non-Christmas books (dotted lines) between June 2021 and December 2024. The Christmas book subset exhibits pronounced seasonal spikes in both local and global drift during November--December each year, while the non-Christmas subset closely tracks the overall all-books trend.
}
\label{drift_christmas_subset}
\end{figure*}

\begin{figure*}[h!] %[H]
\centering
\includegraphics[width=0.9\textwidth]{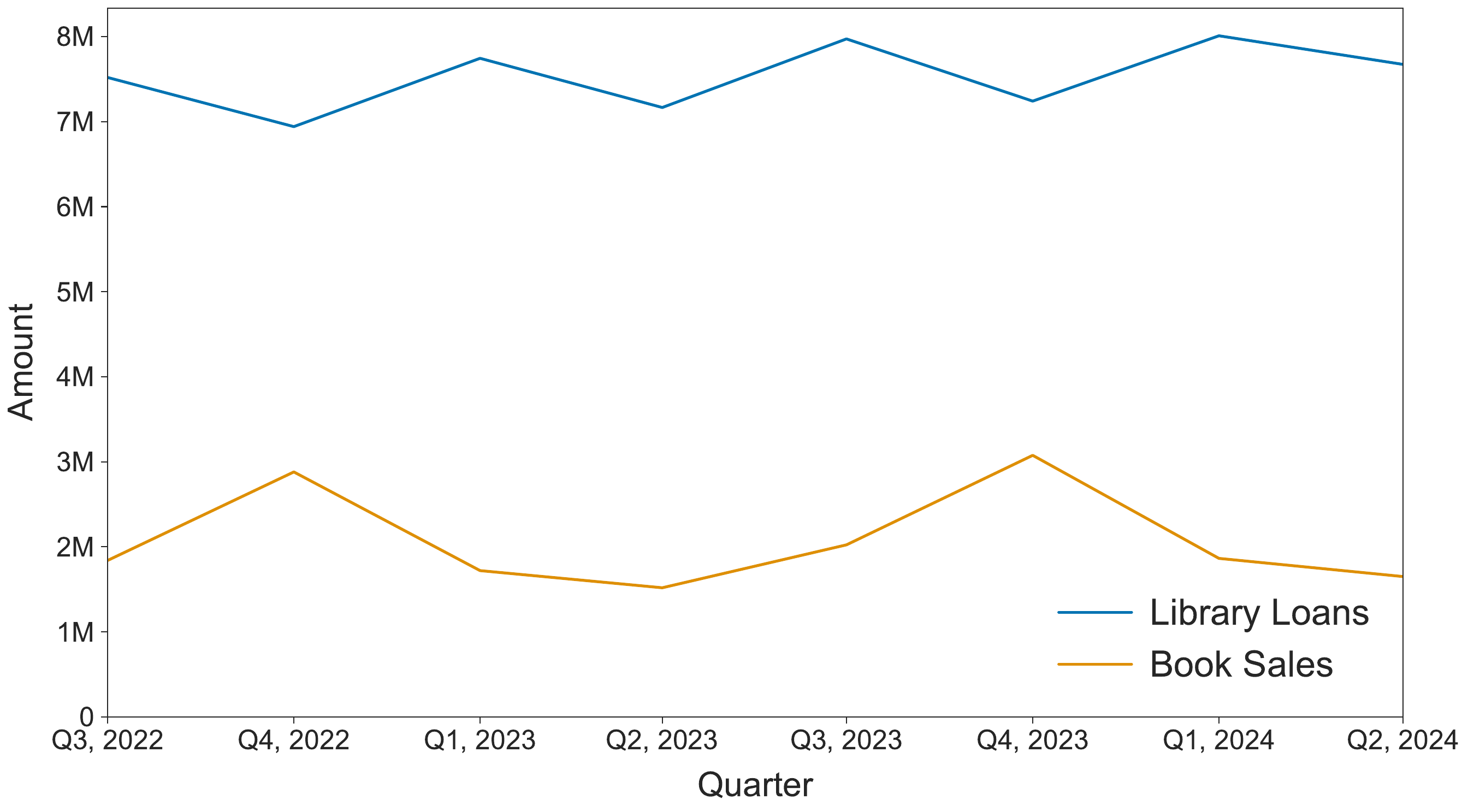}
\caption{
    Comparing the number of loans in our library dataset to the number of books sales in months where both are available. The book sales data stems from Statistics Denmark and contains information on the majority of unit sales of new physical books, e-books, and audiobooks from physical retailers (including the biggest supermarket vendors) as well as webshops~\cite{statisticsdenmark_book_}. Sales of used books are not included.
}
\label{book_sales_loans_comparison}
\end{figure*}

\begin{figure*}[h!] %[H]
\centering
\includegraphics[width=0.9\textwidth]{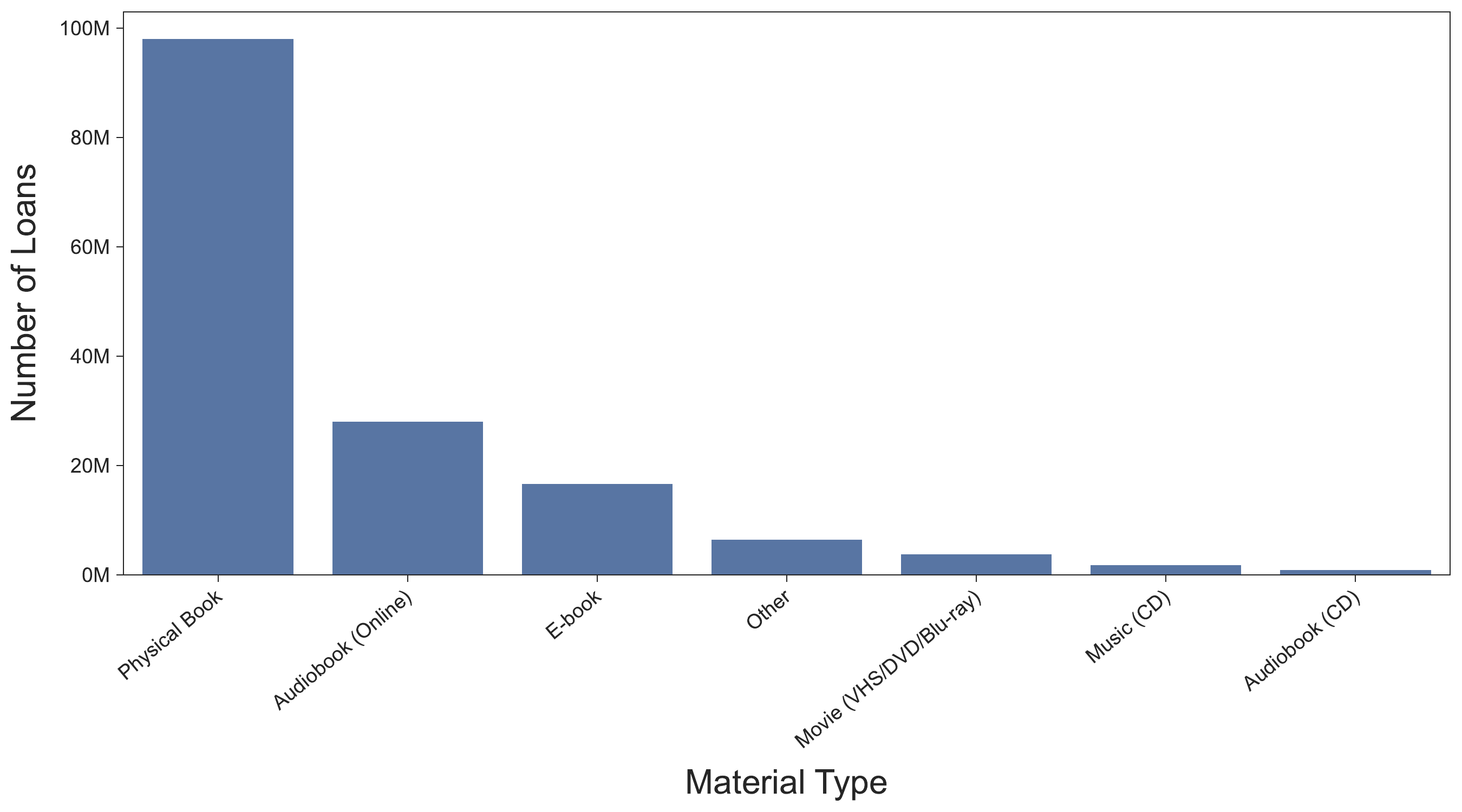}
\caption{
    The number of loans for the different types of materials that can be loaned from Danish public libraries from 2020 to 2024. Materials in the "Other" category feature predominantly magazines but also include games, sheet music, and some more obscure material types. 
}
\label{material_type_distribution}
\end{figure*}

\begin{figure*}[h!] %[H]
\centering
\includegraphics[width=0.8\textwidth]{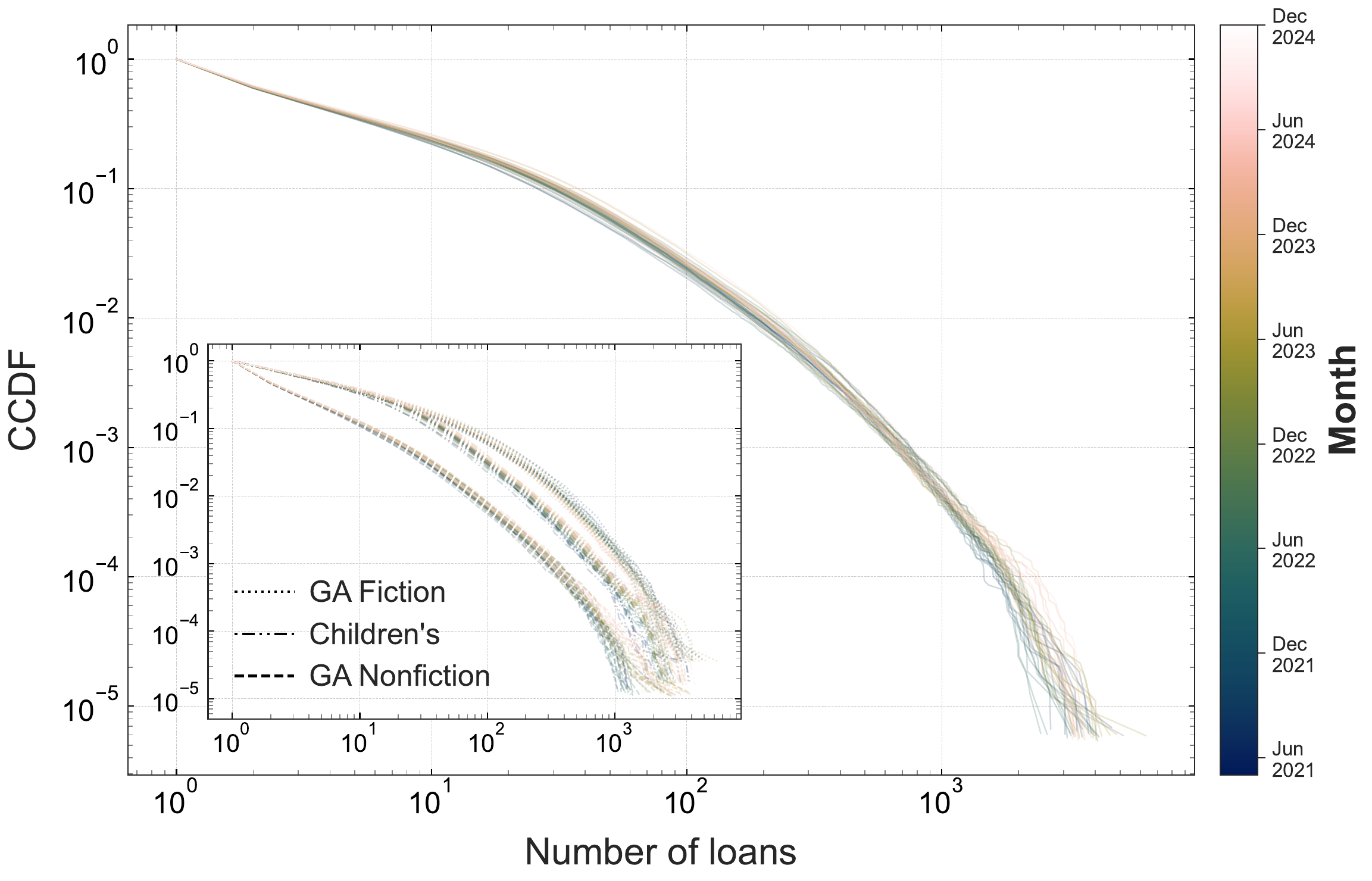}
\caption{
    Complementary cumulative distribution function (CCDF) of the number of book loans for all months after April 2021. The inset shows the CCDFs for three main categories of books.
} 
\label{book_popularity_distributions_months}
\end{figure*}

\begin{figure*}[h!] %[H]
\centering
\includegraphics[width=1.0\textwidth]{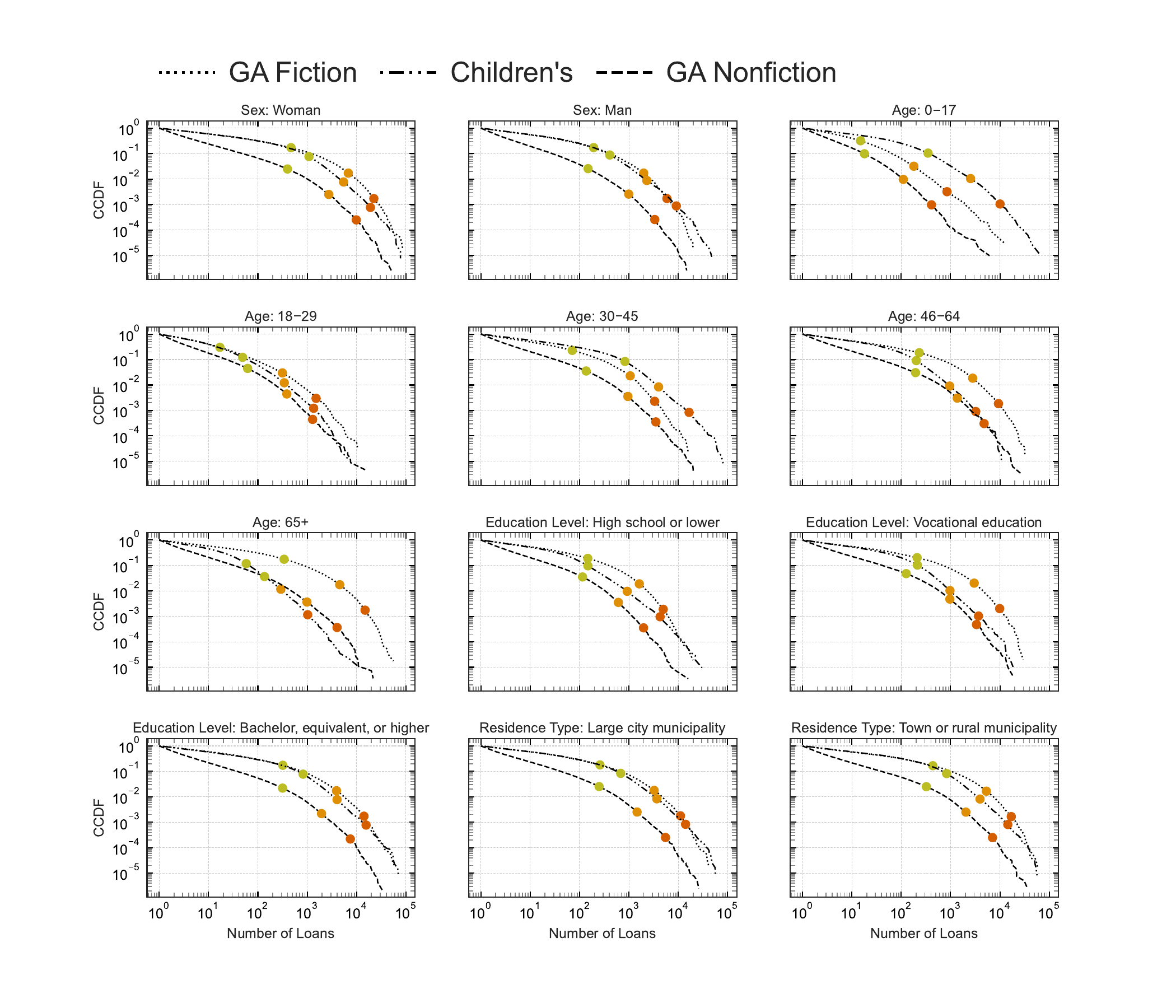}
\caption{
    Complementary cumulative distribution function (CCDF) of the number of book loans over all months for each combination of the three main book categories and loaner subgroups. Green dots mark the top 10,000th most popular book, orange dots mark the top 1,000th most popular book, and red dots mark the 100th most popular book.
} 
\label{book_popularity_distributions_loaner_groups}
\end{figure*}

\begin{figure*}[h!] %[H]
\centering
\includegraphics[width=1.0\textwidth]{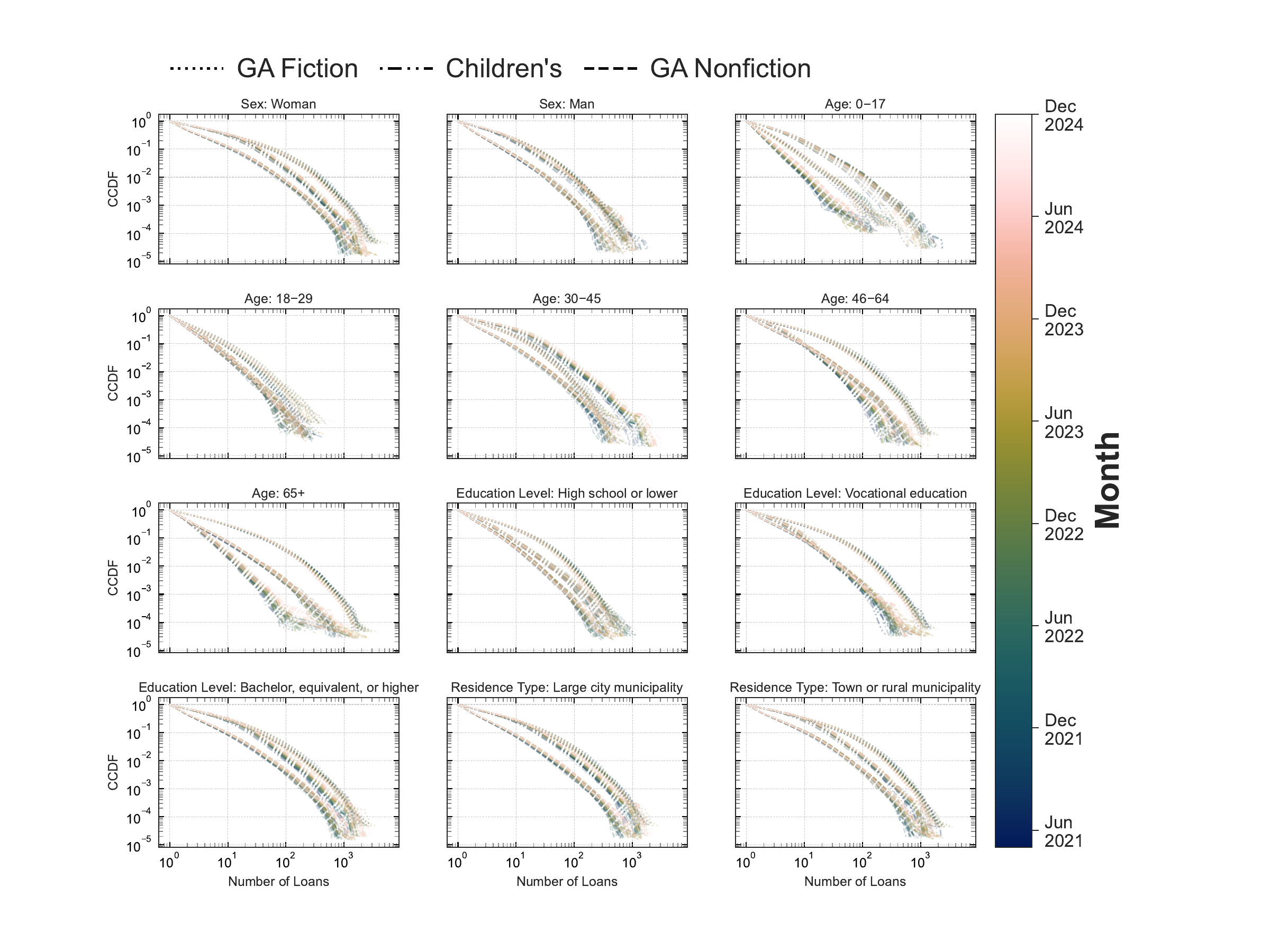}
\caption{
    Complementary cumulative distribution function (CCDF) of the number of book loans for all months after April 2021 for each combination of the three main book categories and loaner subgroups.
} 
\label{book_popularity_distributions_loaner_groups_months}
\end{figure*}

\begin{figure*}[h!] %[H]
\centering
\includegraphics[width=0.9\textwidth]{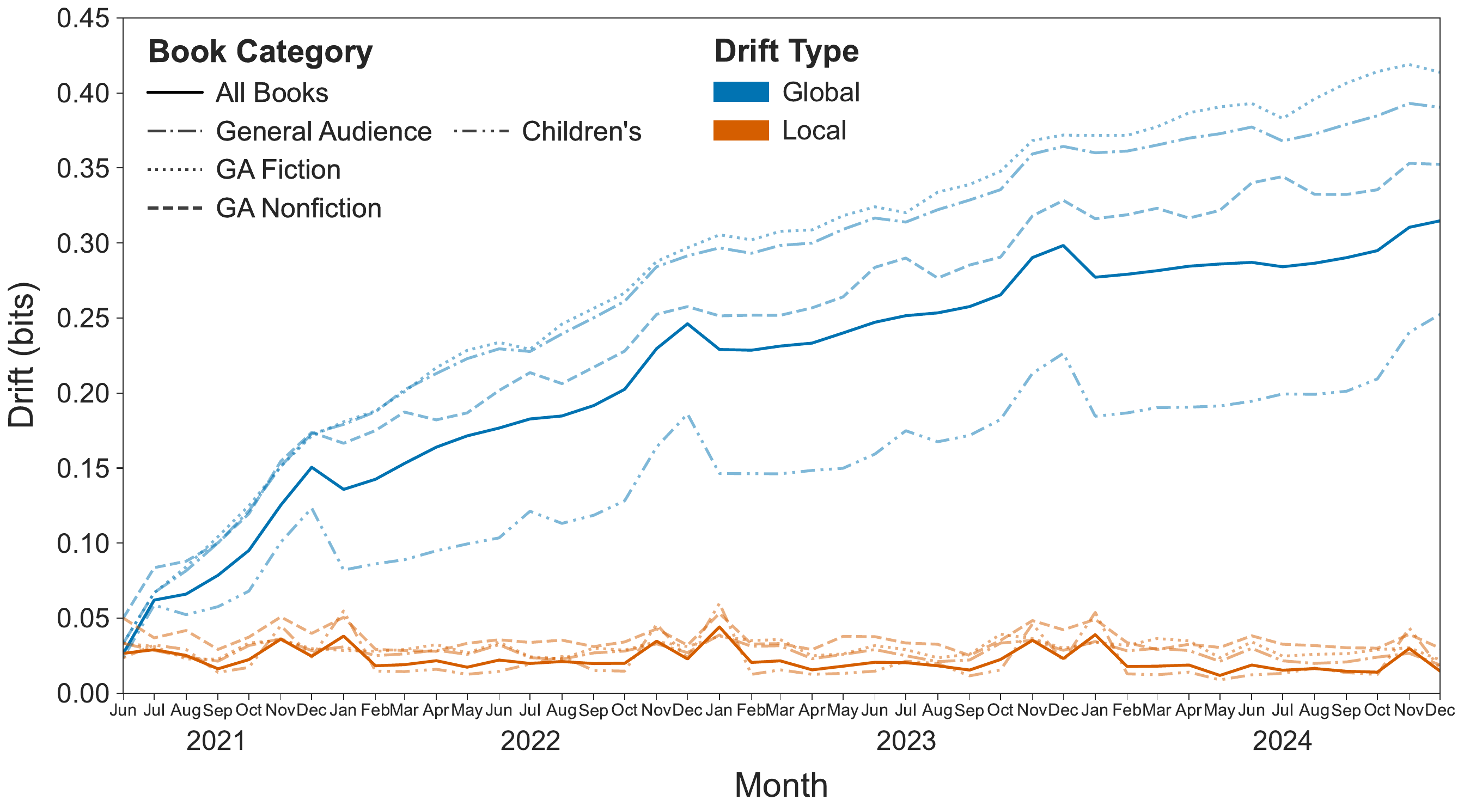}
\caption{
    Local and global drifts in consumer attention for the top 10,000 most popular books across book types, using montly bins.
} 
\label{drifts_top10k}
\end{figure*}

\begin{figure*}[h!] %[H]
\centering
\includegraphics[width=0.9\textwidth]{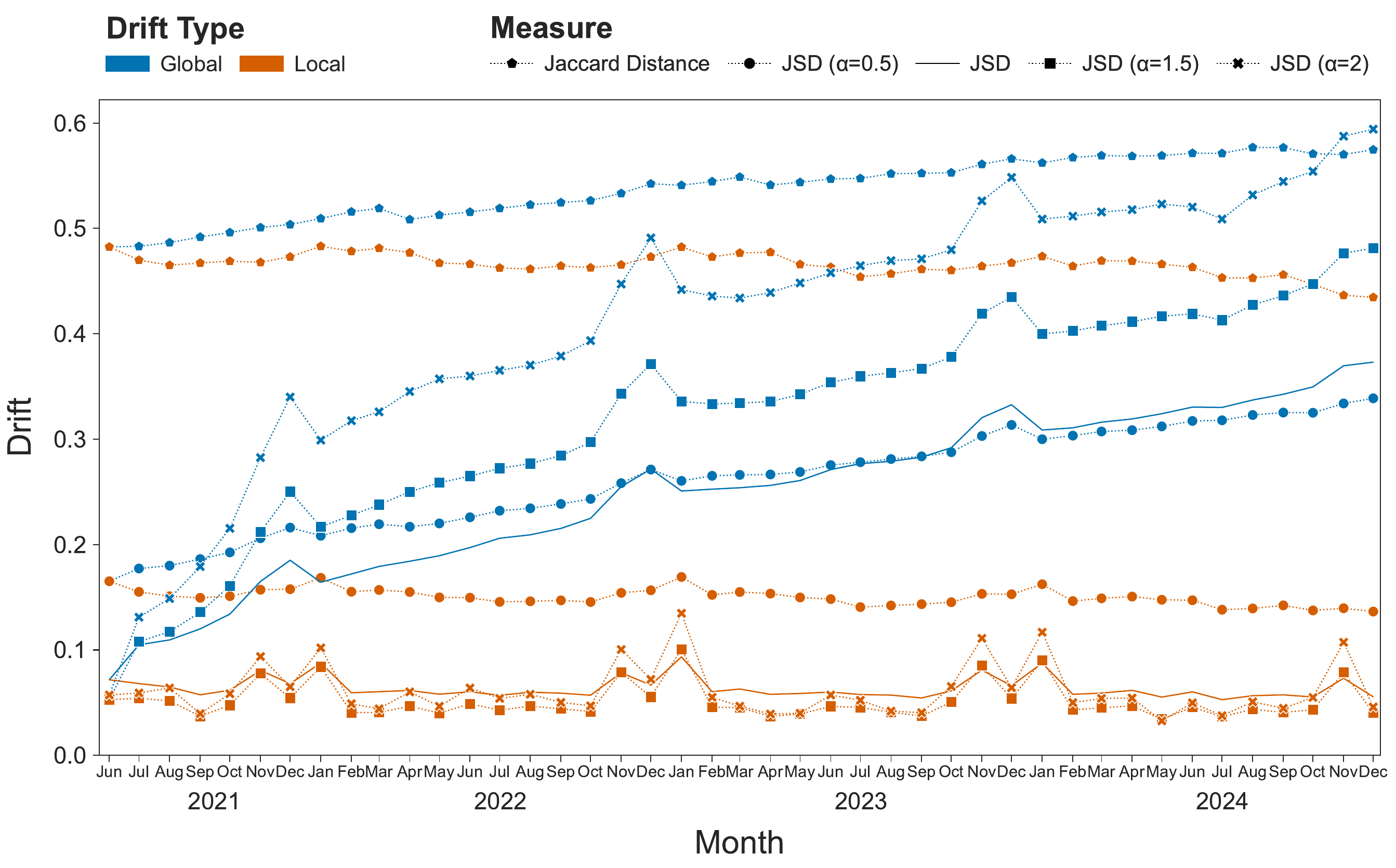}
\caption{
    Measures of drift using the normalized version of the Generalized Jensen-Shannon Divergence \emph{across all books}, varying $\alpha$ to emphasize changes to books with high and low popularity.
} 
\label{drifts_alternative_measures_all_books}
\end{figure*}

\begin{figure*}[h!] %[H]
\centering
\includegraphics[width=0.9\textwidth]{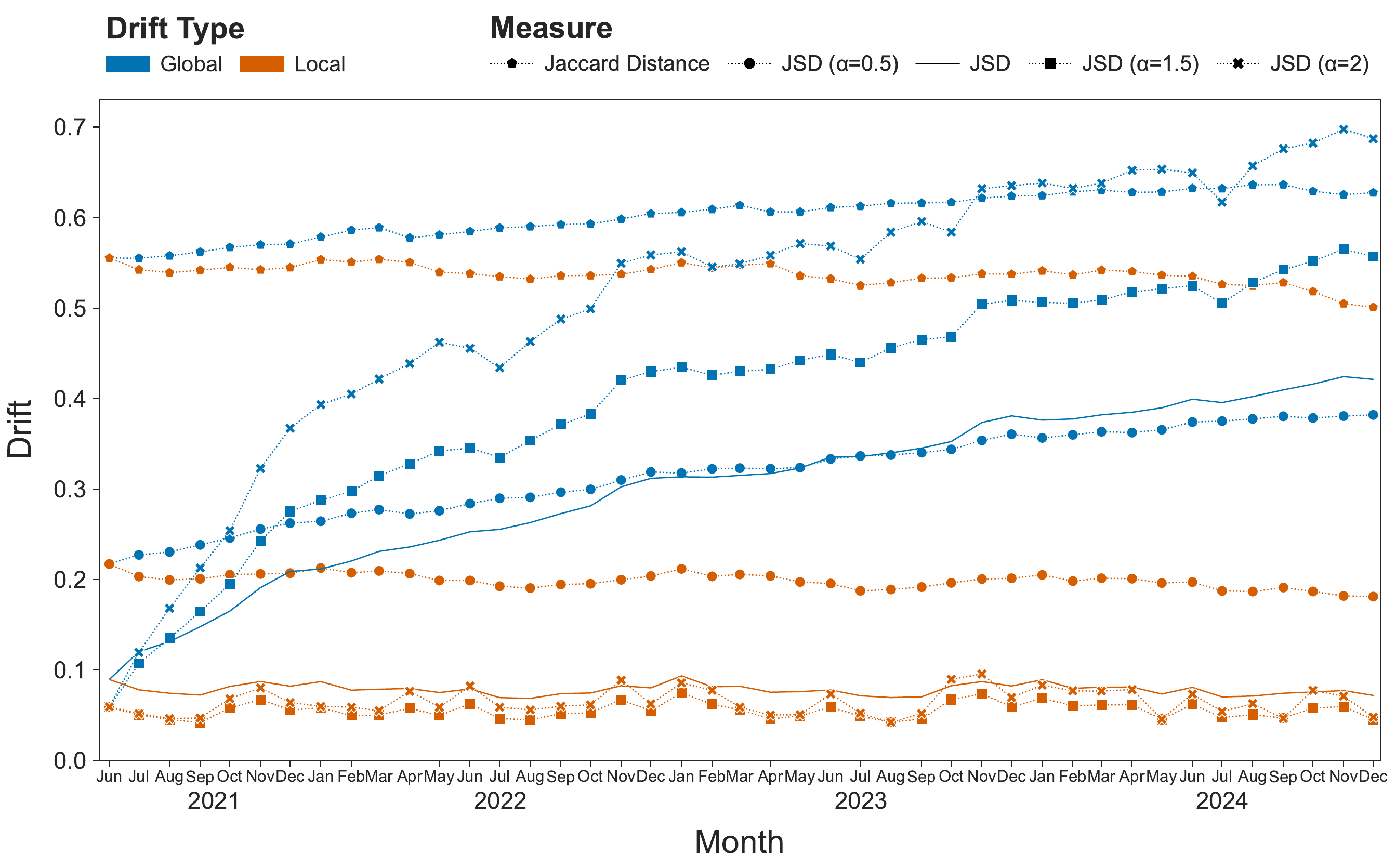}
\caption{
    Measures of drift using the normalized version of the Generalized Jensen-Shannon Divergence \emph{general audience books}, varying $\alpha$ to emphasize changes to books with high and low popularity.
} 
\label{drifts_alternative_measures_ga}
\end{figure*}

\begin{figure*}[h!] %[H]
\centering
\includegraphics[width=0.9\textwidth]{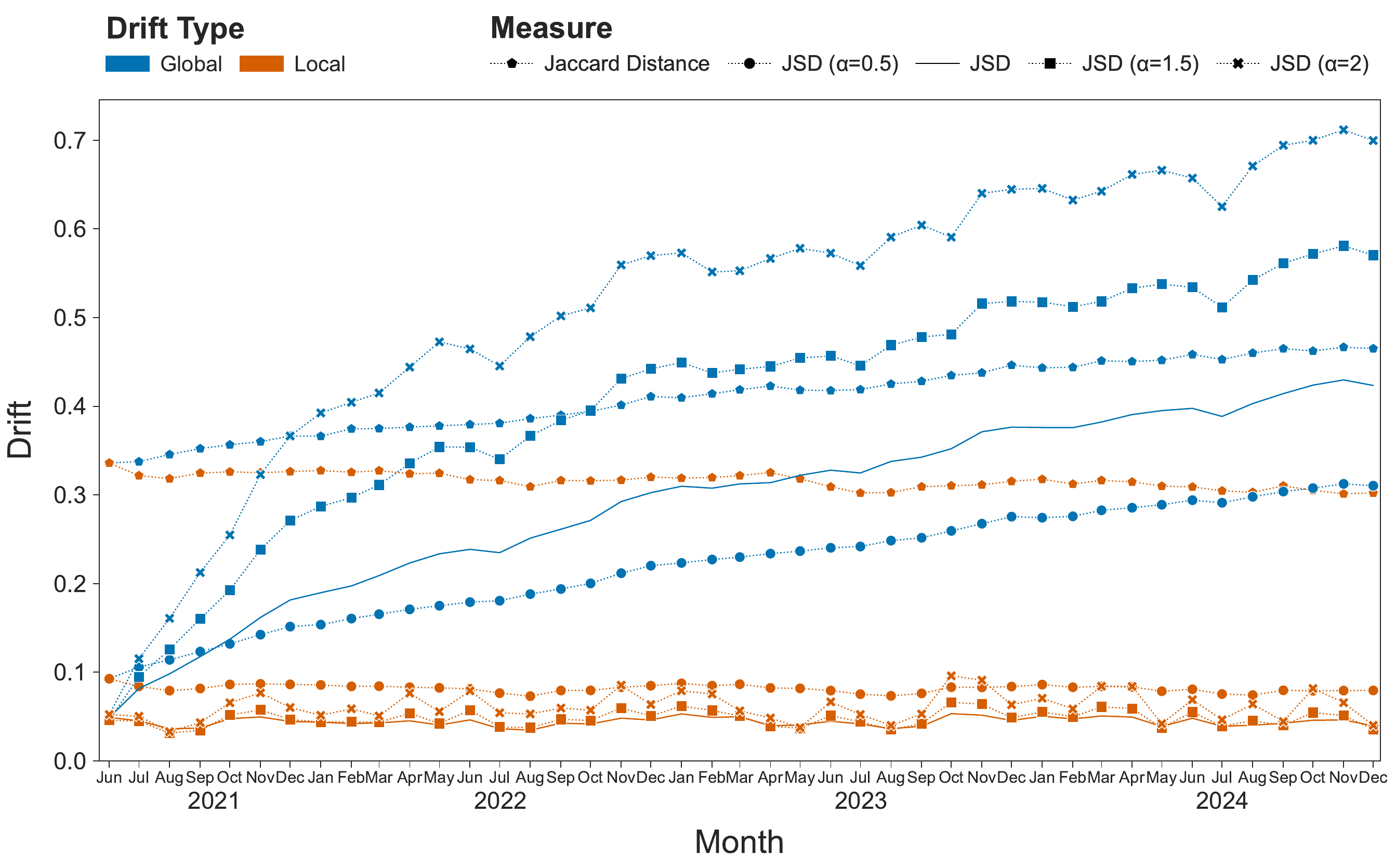}
\caption{
    Measures of drift using the normalized version of the Generalized Jensen-Shannon Divergence for \emph{general audience fiction books}, varying $\alpha$ to emphasize changes to books with high and low popularity.
    } 
\label{drifts_alternative_measures_ga_fic}
\end{figure*}

\begin{figure*}[h!] %[H]
\centering
\includegraphics[width=0.9\textwidth]{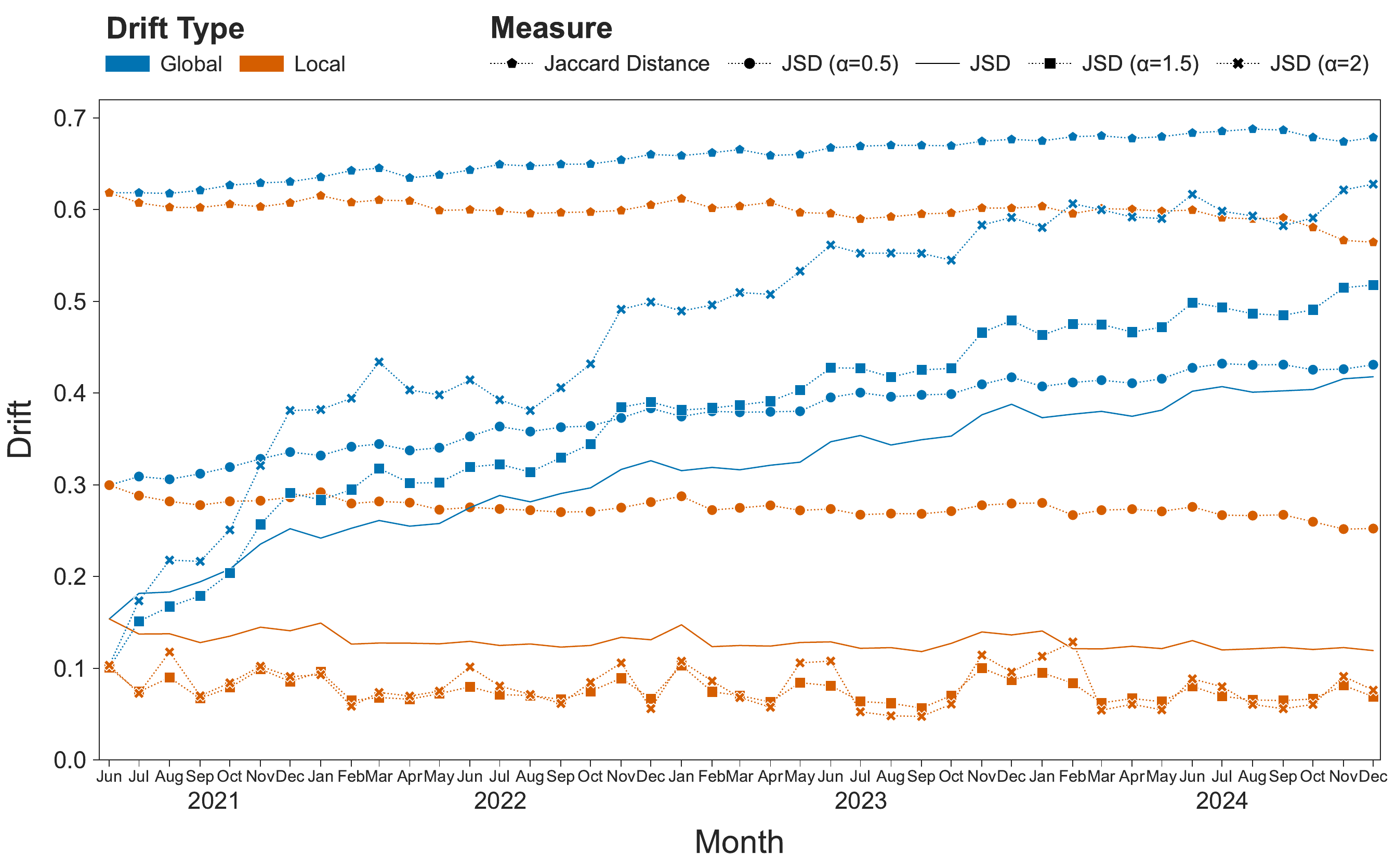}
\caption{
    Measures of drift using the normalized version of the Generalized Jensen-Shannon Divergence for \emph{general audience nonfiction books}, varying $\alpha$ to emphasize changes to books with high and low popularity.
} 
\label{drifts_alternative_measures_ga_nonfic}
\end{figure*}

\begin{figure*}[h!] %[H]
\centering
\includegraphics[width=0.9\textwidth]{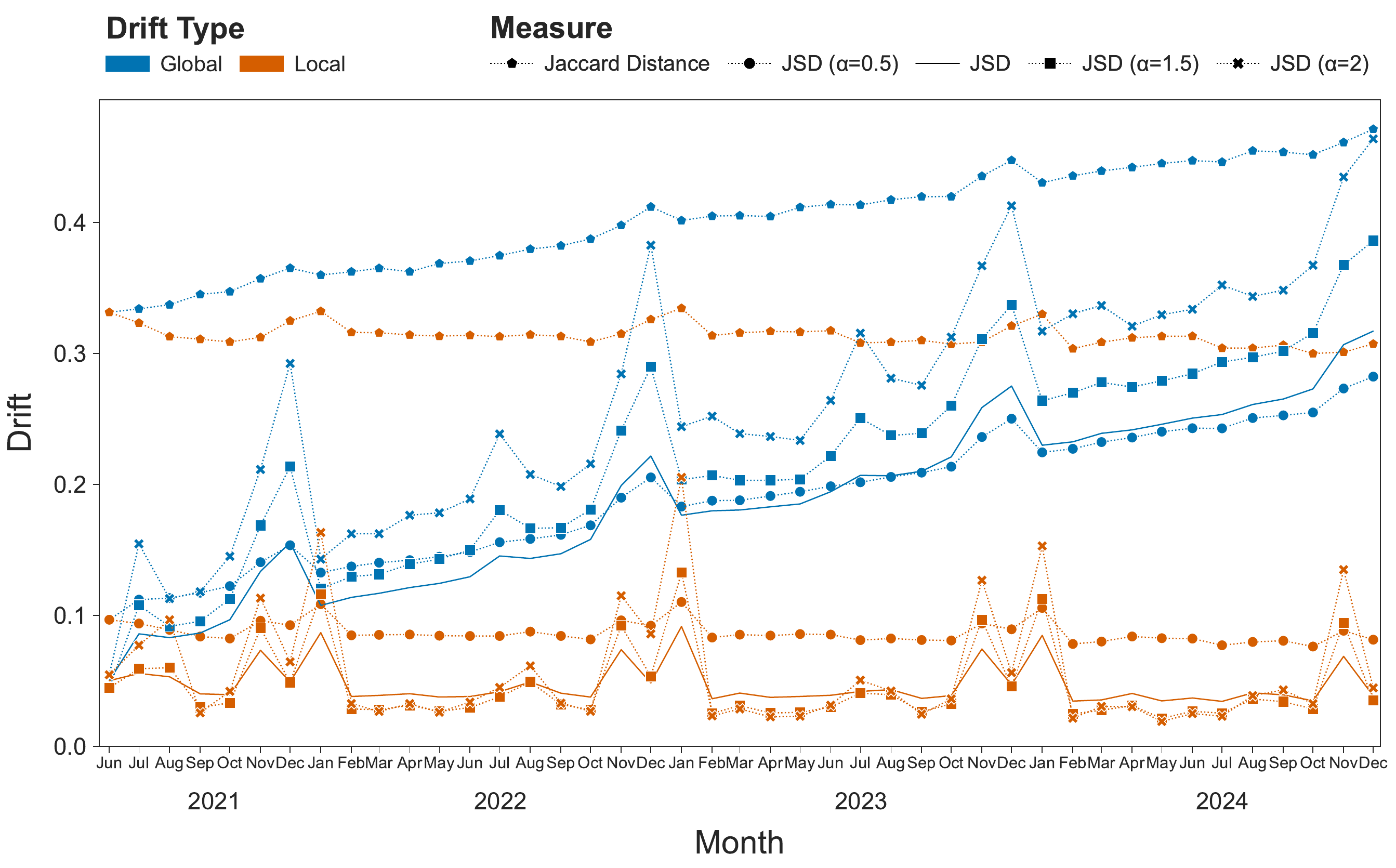}
\caption{
    Measures of drift using the normalized version of the Generalized Jensen-Shannon Divergence for \emph{children's books}, varying $\alpha$ to emphasize changes to books with high and low popularity.
} 
\label{drifts_alternative_measures_children}
\end{figure*}

\begin{figure*}[h!] %[H]
\centering
\includegraphics[width=0.9\textwidth]{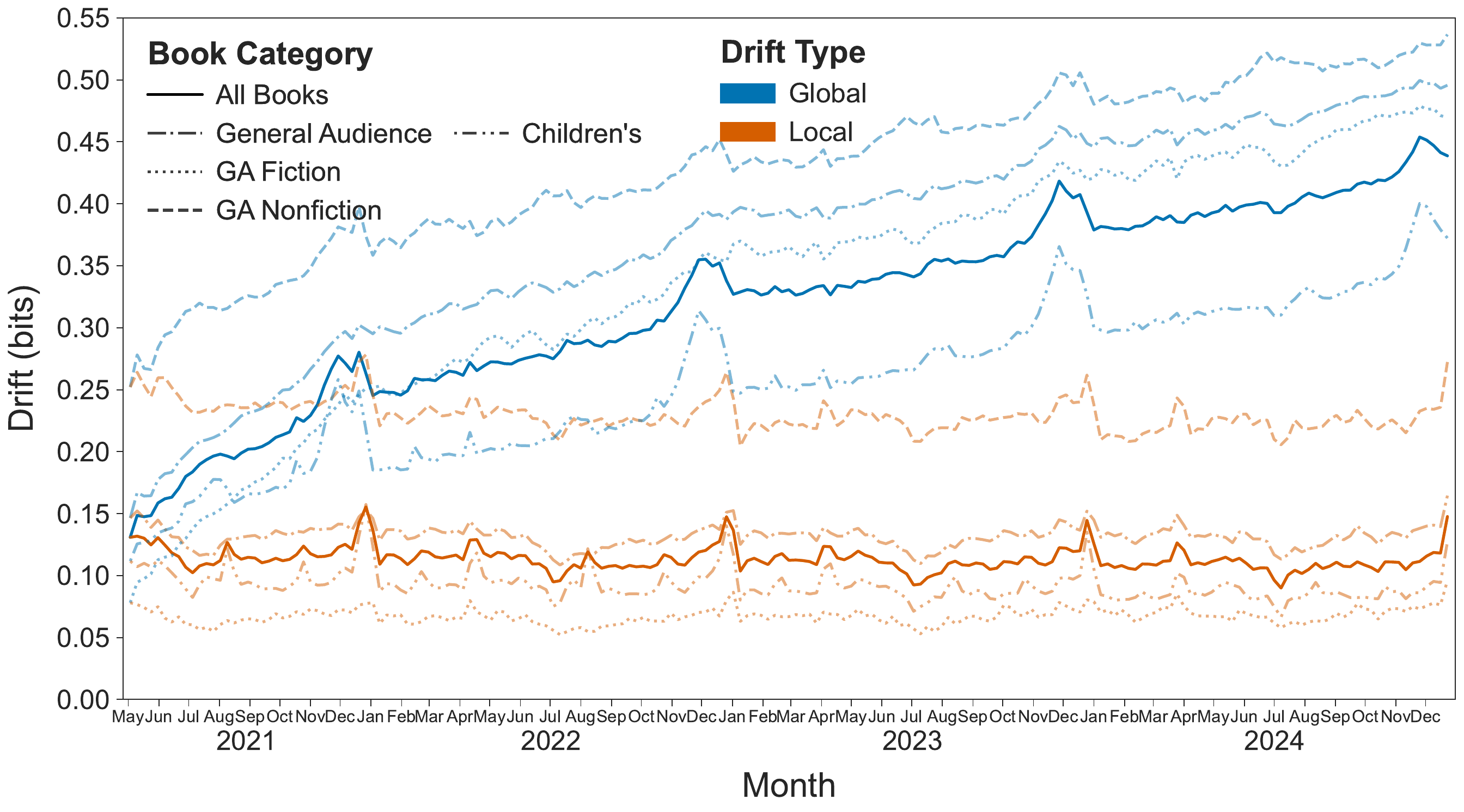}
\caption{
    Local and global drifts in consumer attention, using weekly bins to construct distributions of popularity. We use the week starting on Monday, 26/04/2021, as the baseline week to compute global drifts.
} 
\label{drifts_week}
\end{figure*}

\begin{figure*}[h!] %[H]
\centering
\includegraphics[width=0.9\textwidth]{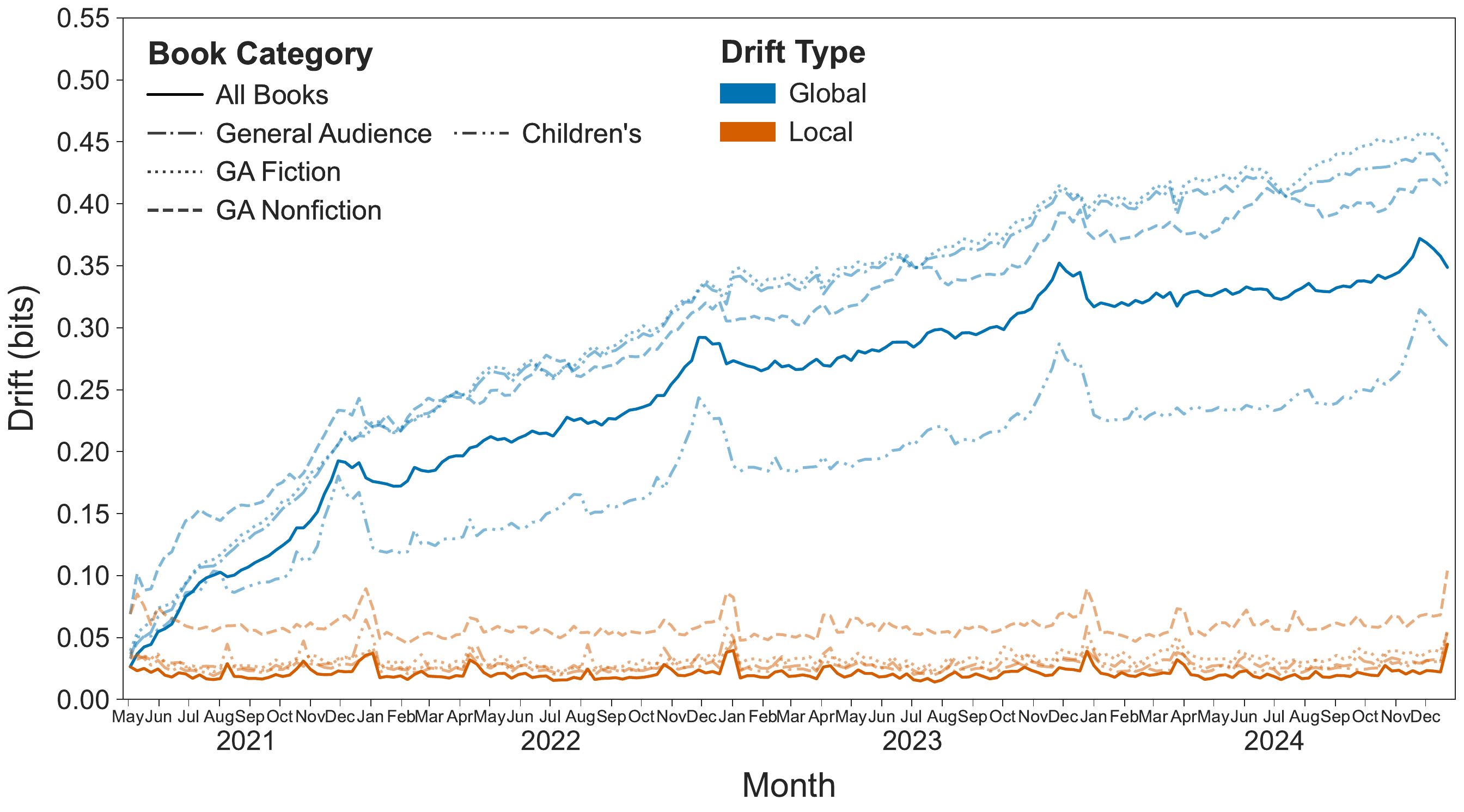}
\caption{
    Local and global drifts in consumer attention for the top 10,000 most popular books, using weekly bins to construct distributions of popularity. We use the week starting on Monday, 26/04/2021, as the baseline week to compute global drifts.
} 
\label{drifts_week_top10k}
\end{figure*}

\begin{figure*}[h!] %[H]
\centering
\includegraphics[width=0.9\textwidth]{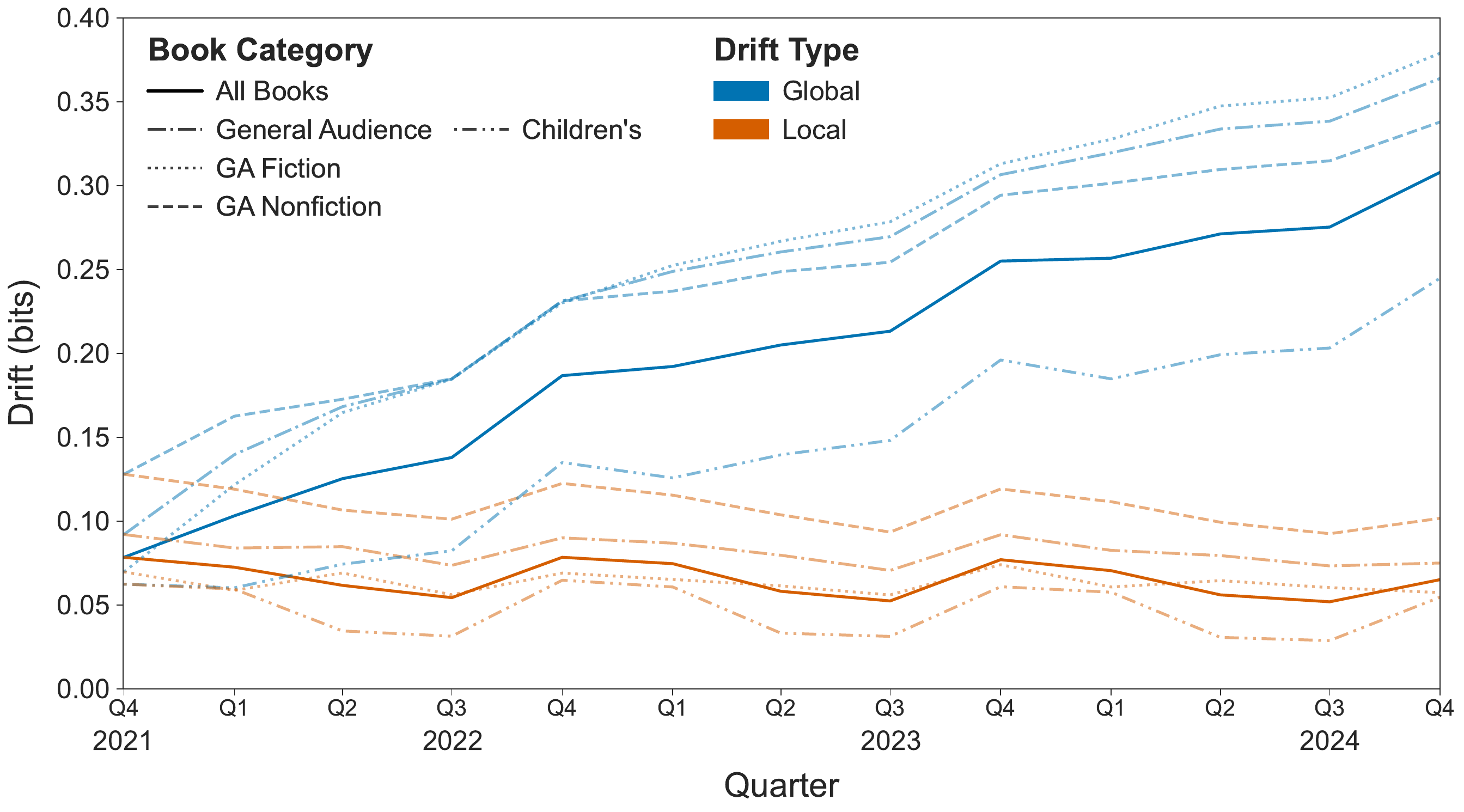}
\caption{
    Local and global drifts in consumer attention, using quarterly bins to construct distributions of popularity. We use the third quarter of 2021 (from 01/07-2021 to 30/09-2021) as the baseline to compute global drifts.
}
\label{drifts_quarter}
\end{figure*}

\begin{figure*}[h!] %[H]
\centering
\includegraphics[width=0.9\textwidth]{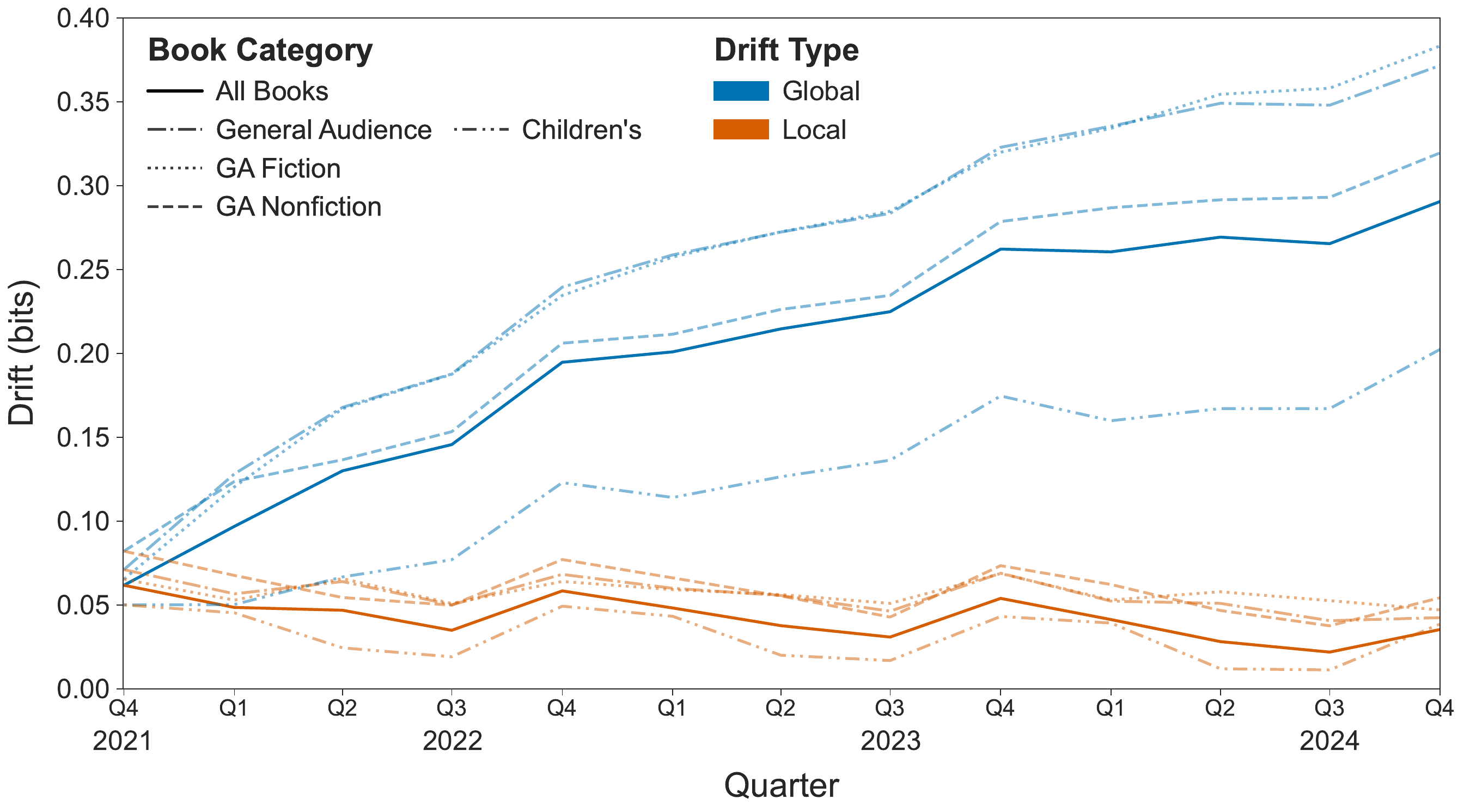}
\caption{
    Local and global drifts in consumer attention for the top 10,000 most popular books, using quarterly bins to construct distributions of popularity. We use the third quarter of 2021 (from 01/07-2021 to 30/09-2021) as the baseline to compute global drifts.
}
\label{drifts_quarter_top10k}
\end{figure*}

\begin{figure*}[h!] %[H]
\centering
\includegraphics[width=0.9\textwidth]{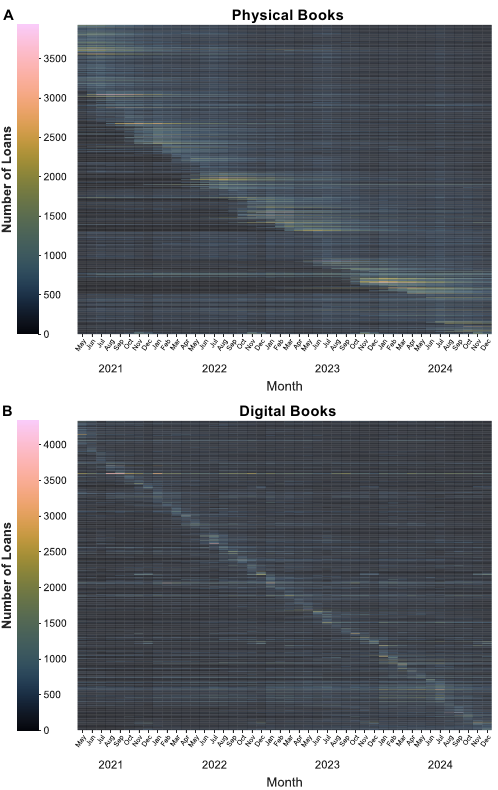}
\caption{
    Monthly popularity for the 1,000 most popular books physical books (\textbf{A}) and 1,000 most popular digital books (\tewxtbf{B}) between May 2021 and December 2024, ordered by the month where their popularity peaks. 
}
\label{popular_overall_counts_top1k_phys-v-dig}
\end{figure*}

\begin{figure*}[h!] %[H]
\centering
\includegraphics[width=0.9\textwidth]{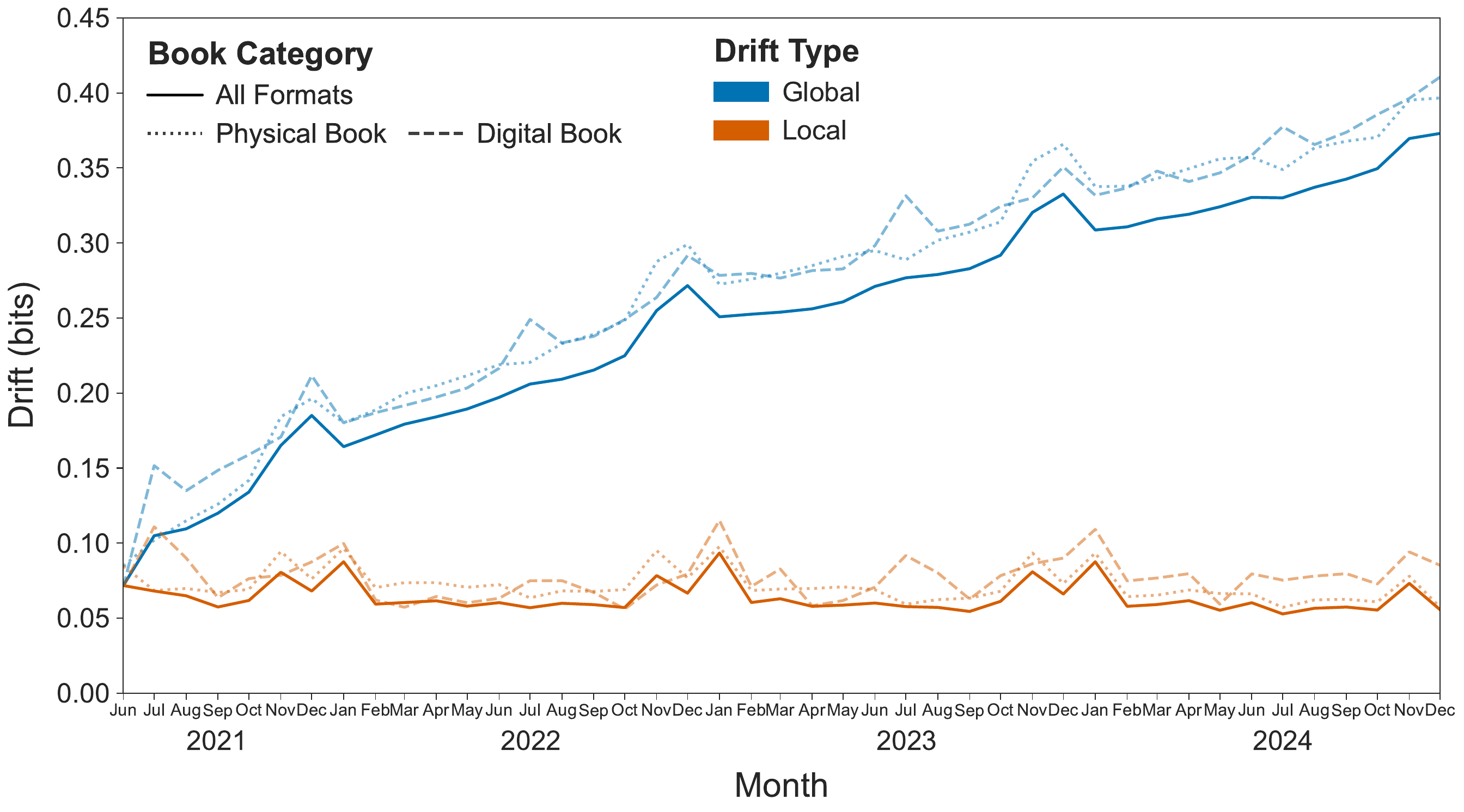}
\caption{
    Local and global drifts in collective attention, computed separately for physical books and digital formats (e-books and audiobooks). Patterns are nearly identical across formats.
}
\label{drifts_phys-v-dig}
\end{figure*}

\begin{figure*}[h!] %[H]
\centering
\includegraphics[width=0.9\textwidth]{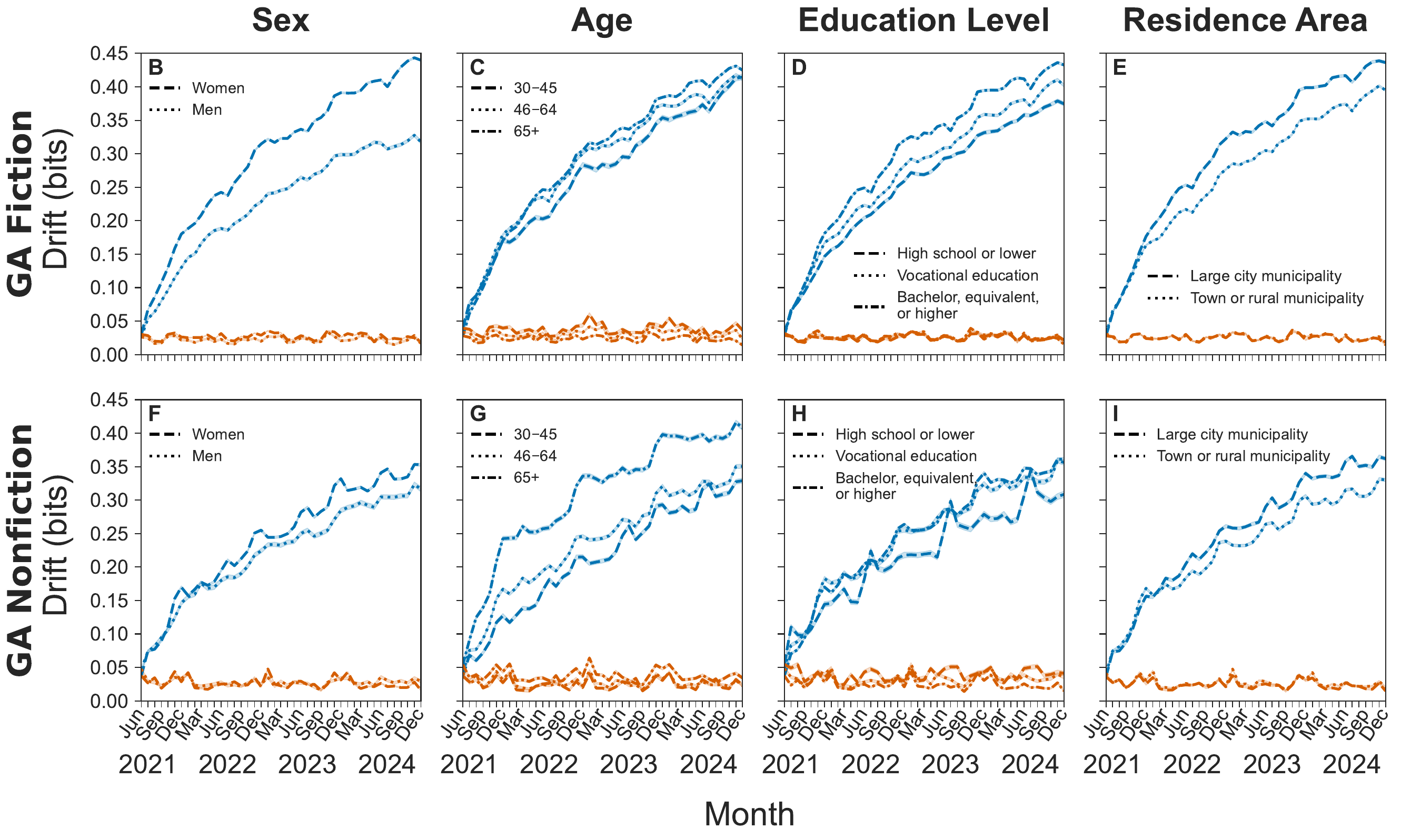}
\caption{
    Local and global drifts for the top 10,000 most popular general audience fiction books (\textbf{B-E}) and top 10,000 most popular general audience nonfiction books (\textbf{F-I}) within different groups of the population. Shaded areas around the lines (which are barely visible) show $2 \sigma$ bootstrap errors. 
} 
\label{drifts_subgroups_ga_fic_nonfic}
\end{figure*}

\begin{figure*}[h!] %[H]
\centering
\includegraphics[width=0.8\textwidth]{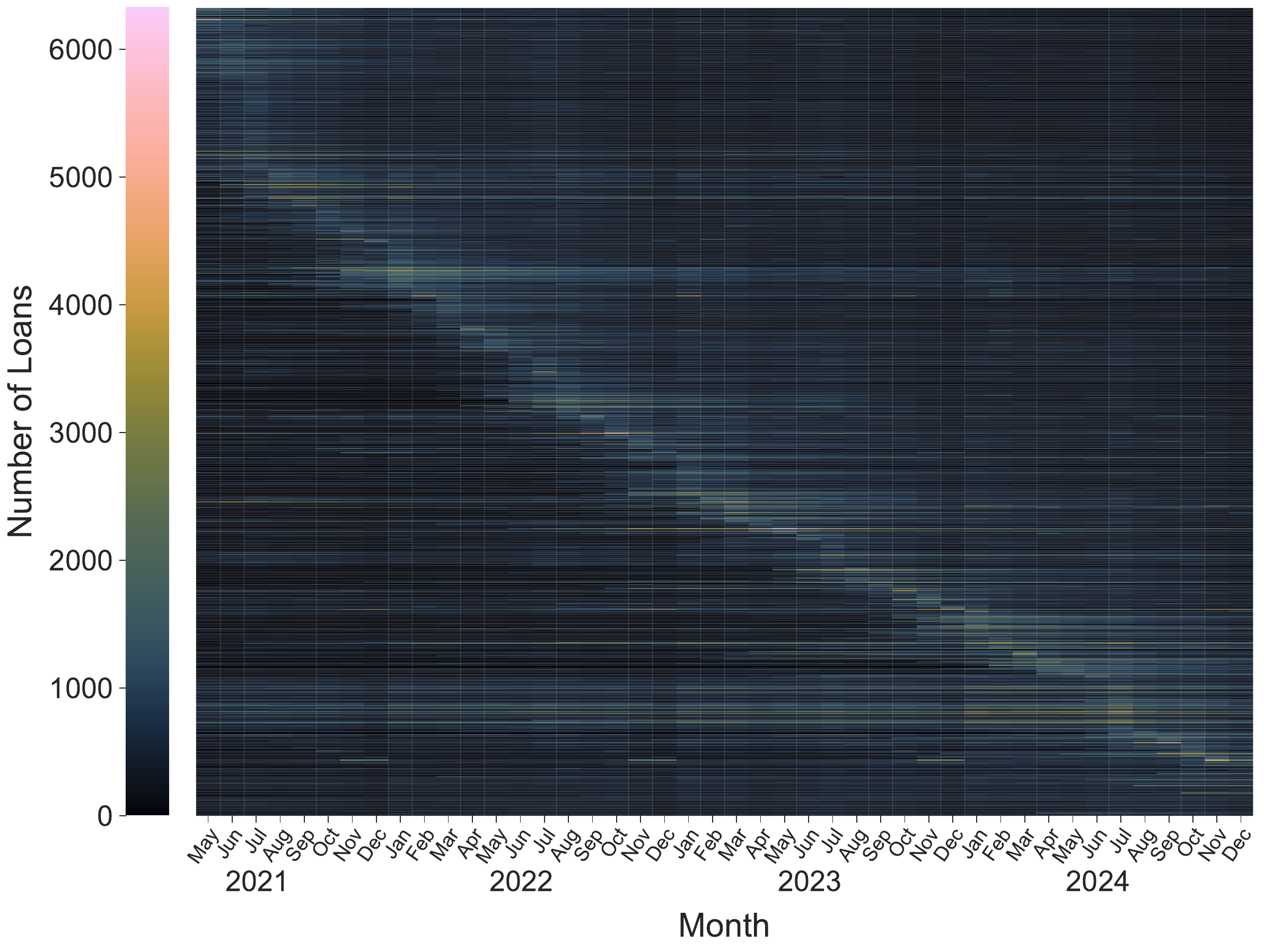}
\caption{
    Monthly popularity for the 1,000 books with the highest total number of loans between May 2021 and 2023, ordered by the month where their popularity peaks.
} 
\label{popular_overall_counts}
\end{figure*}

\begin{figure*}[h!] %[H]
\centering
\includegraphics[width=0.8\textwidth]{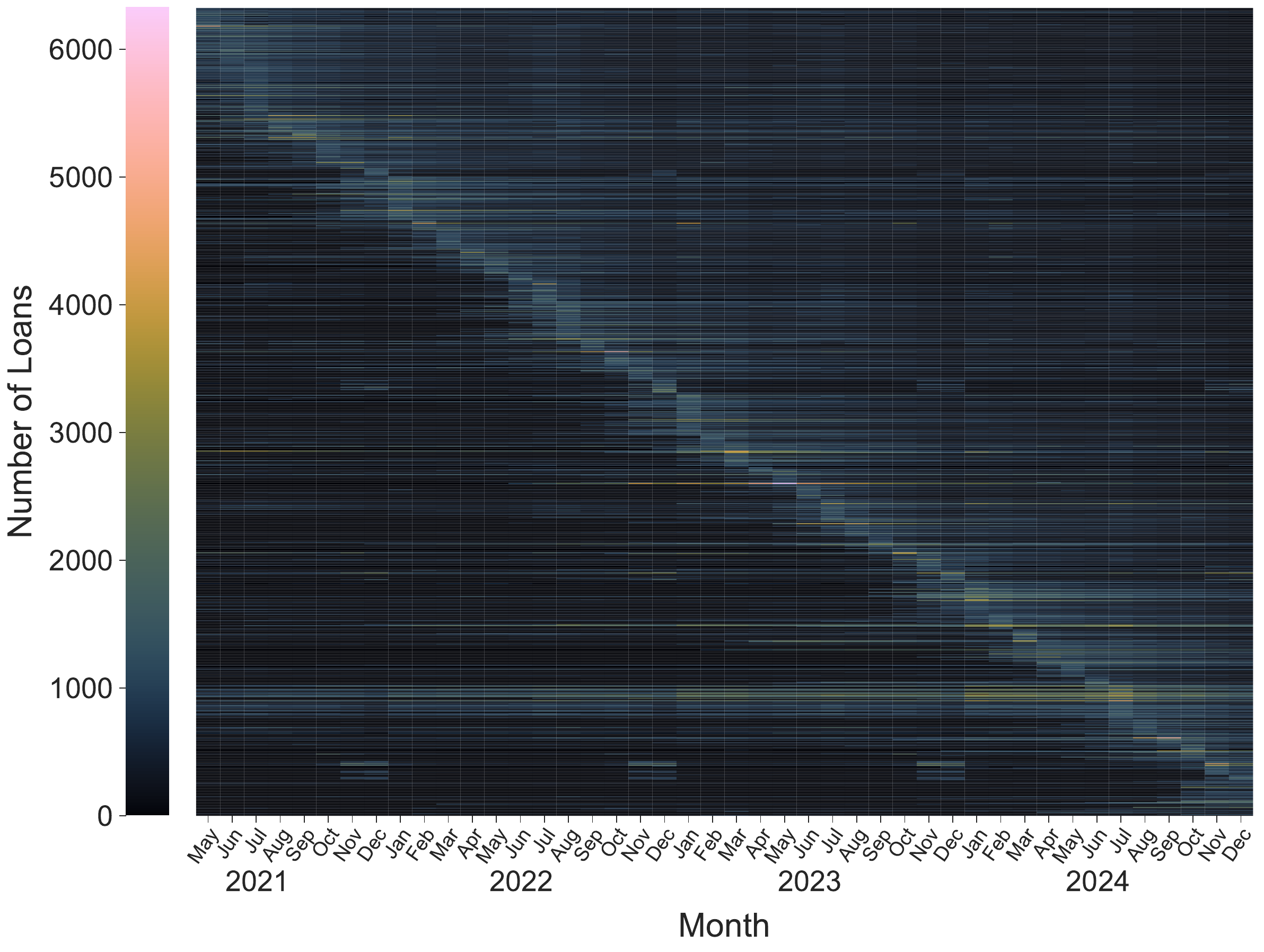}
\caption{
    Monthly popularity for the 1,000 books with the highest peak number of loans in a month between May 2021 and 2023, ordered by the month where their popularity peaks.
} 
\label{peak_monthly_popular_counts}
\end{figure*}

\begin{figure*}[h!] %[H]
\centering
\includegraphics[width=0.8\textwidth]{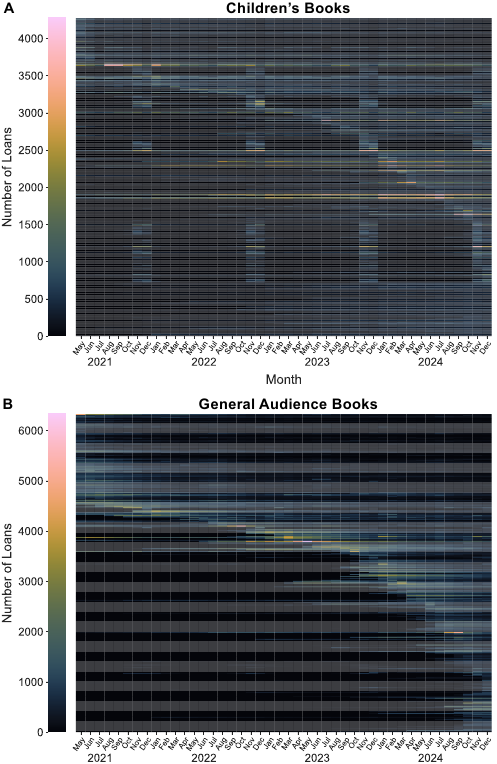}
\caption{
    Monthly popularity of the 1,000 books books with the highest contributions to the global drift in December 2024, shown for children's books (\textbf{A}) and general audience books (\textbf{B}), ordered by the month where their popularity peaks. Children’s books exhibit pronounced recurring seasonal patterns, whereas such recurrence is largely absent among general audience books.
} 
\label{most_contributing_dec2024}
\end{figure*}

\begin{figure*}[h!] %[H]
\centering
\includegraphics[width=1.0\textwidth]{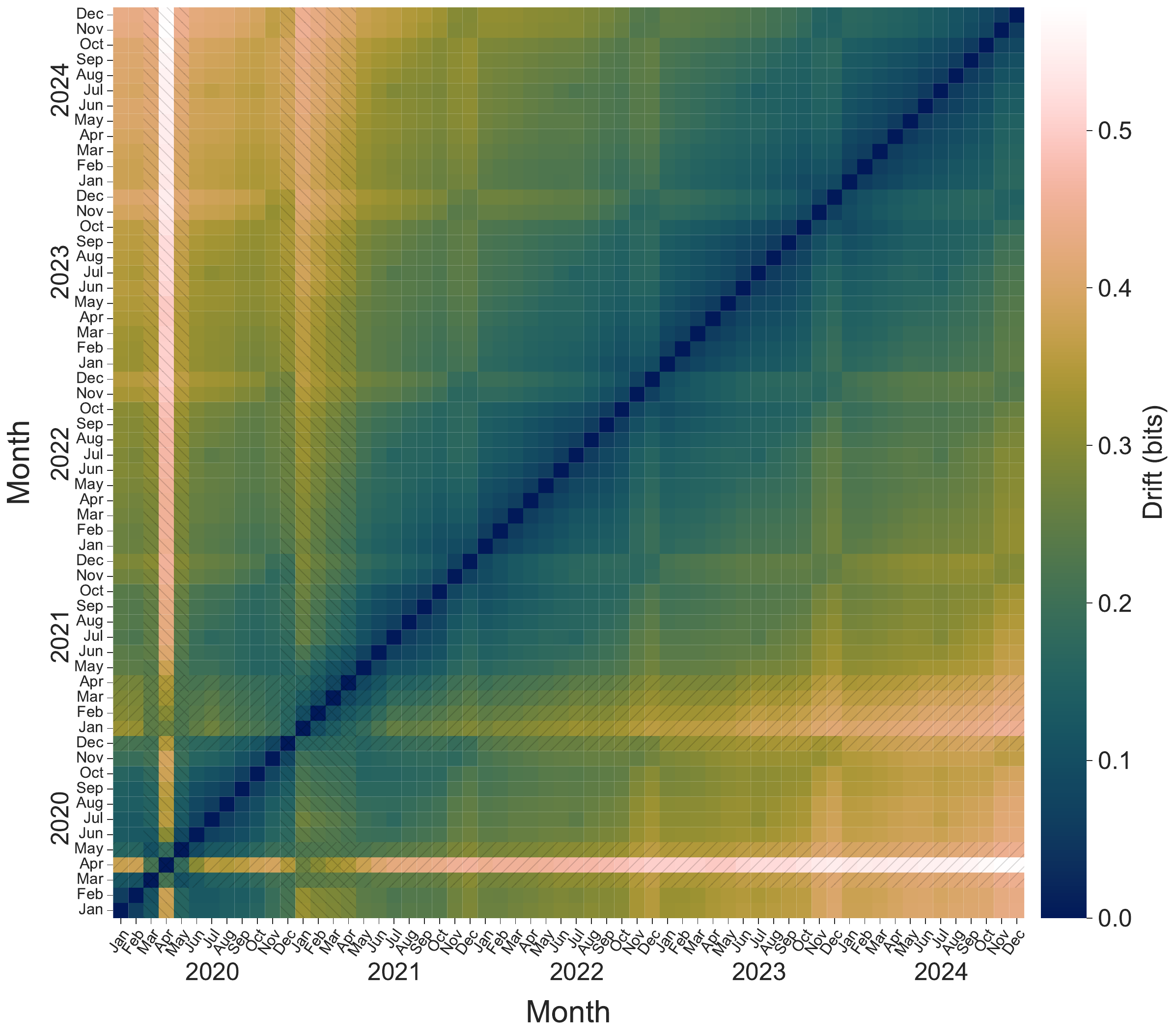}
\caption{
    Drifts in the collective attention for \emph{all books} between all pairs of months in in the dataset. Tiles are marked with a one-way hatch if one month was affected by a COVID-19 lockdown, and a two-way hatch if both months were affected.
} 
\label{drifts_heatmap}
\end{figure*}

\begin{figure*}[h!] %[H]
\centering
\includegraphics[width=1.0\textwidth]{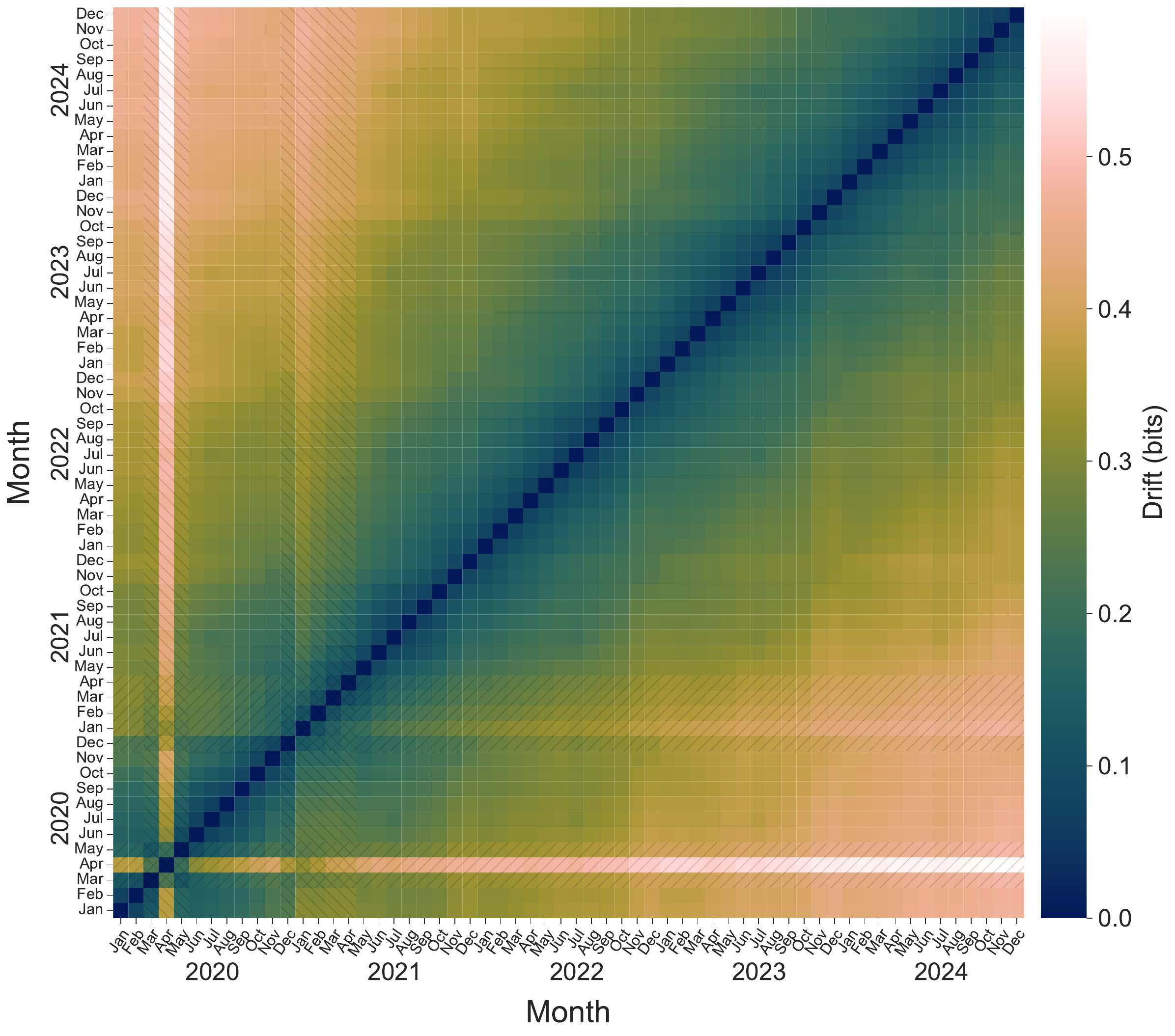}
\caption{
    Drifts in the collective attention for \emph{general audience books} between all pairs of months in in the dataset. Tiles are marked with a one-way hatch if one month was affected by a COVID-19 lockdown, and a two-way hatch if both months were affected.
} 
\label{drifts_heatmap_ga}
\end{figure*}

\begin{figure*}[h!] %[H]
\centering
\includegraphics[width=1.0\textwidth]{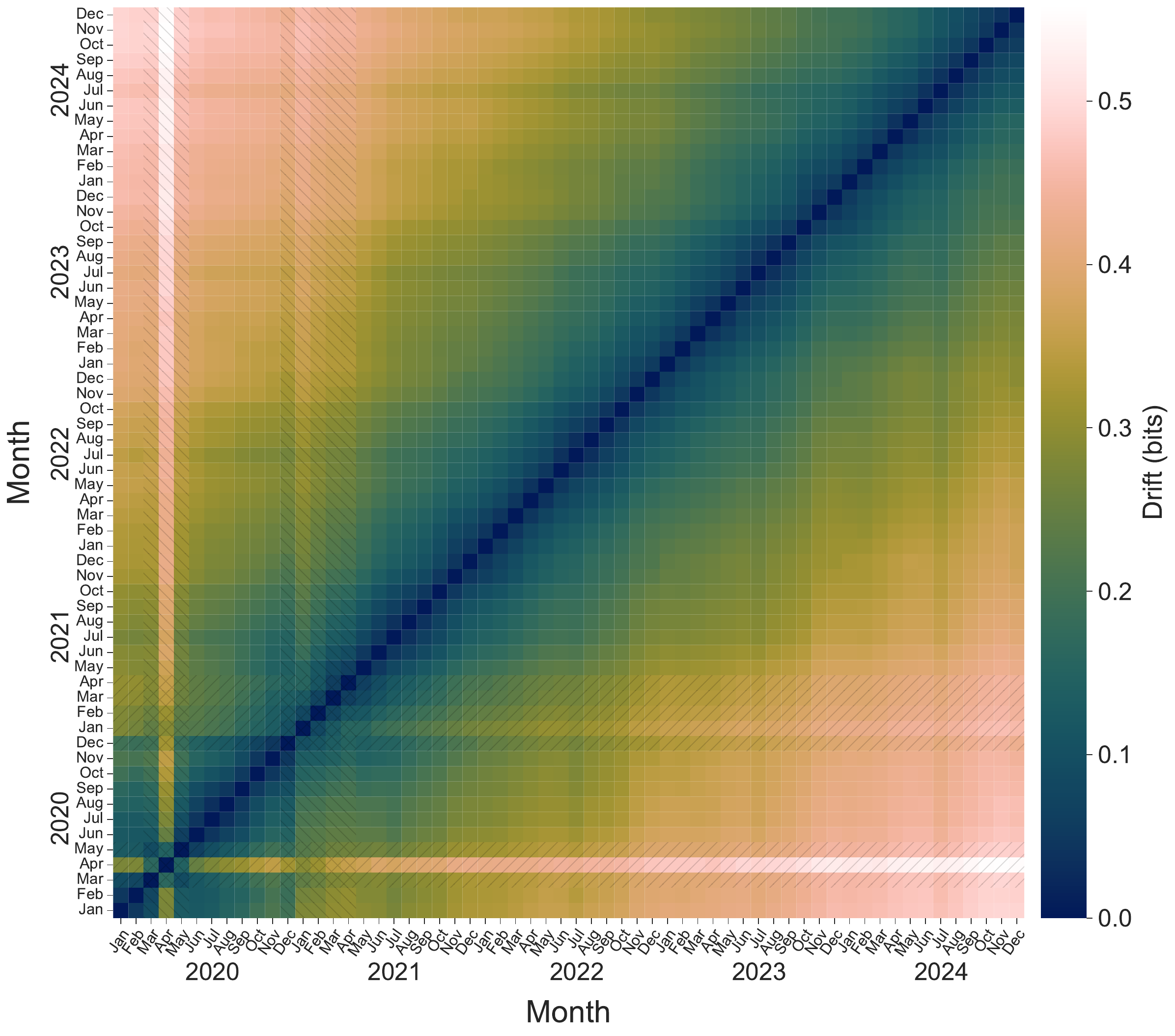}
\caption{
    Drifts in the collective attention for \emph{general audience fiction books} between all pairs of months in in the dataset. Tiles are marked with a one-way hatch if one month was affected by a COVID-19 lockdown, and a two-way hatch if both months were affected.
} 
\label{drifts_heatmap_ga_fic}
\end{figure*}

\begin{figure*}[h!] %[H]
\centering
\includegraphics[width=1.0\textwidth]{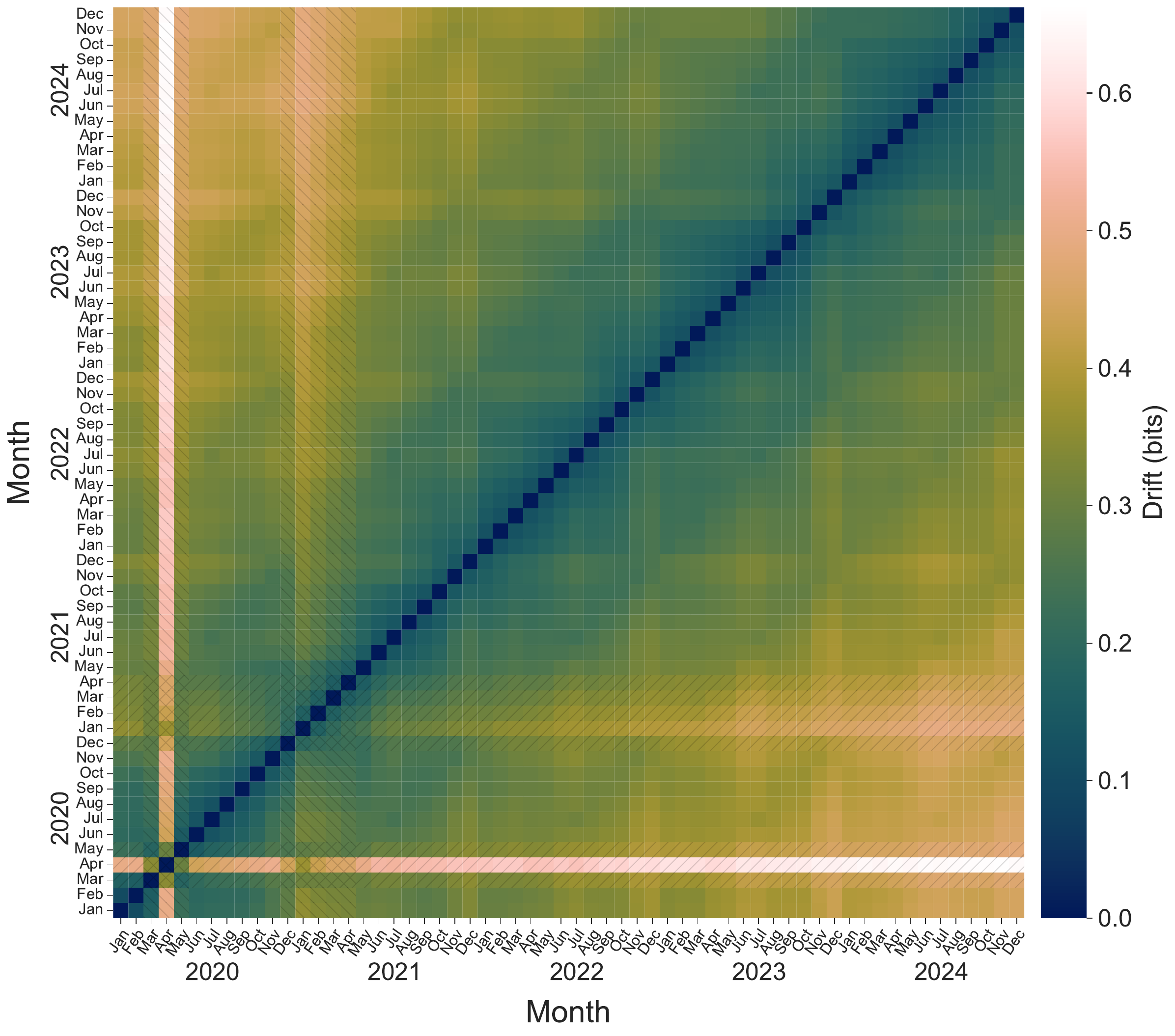}
\caption{
    Drifts in the collective attention for \emph{general audience nonfiction books} between all pairs of months in in the dataset. Tiles are marked with a one-way hatch if one month was affected by a COVID-19 lockdown, and a two-way hatch if both months were affected.
} 
\label{drifts_heatmap_ga_nonfic}
\end{figure*}

\begin{figure*}[h!] %[H]
\centering
\includegraphics[width=1.0\textwidth]{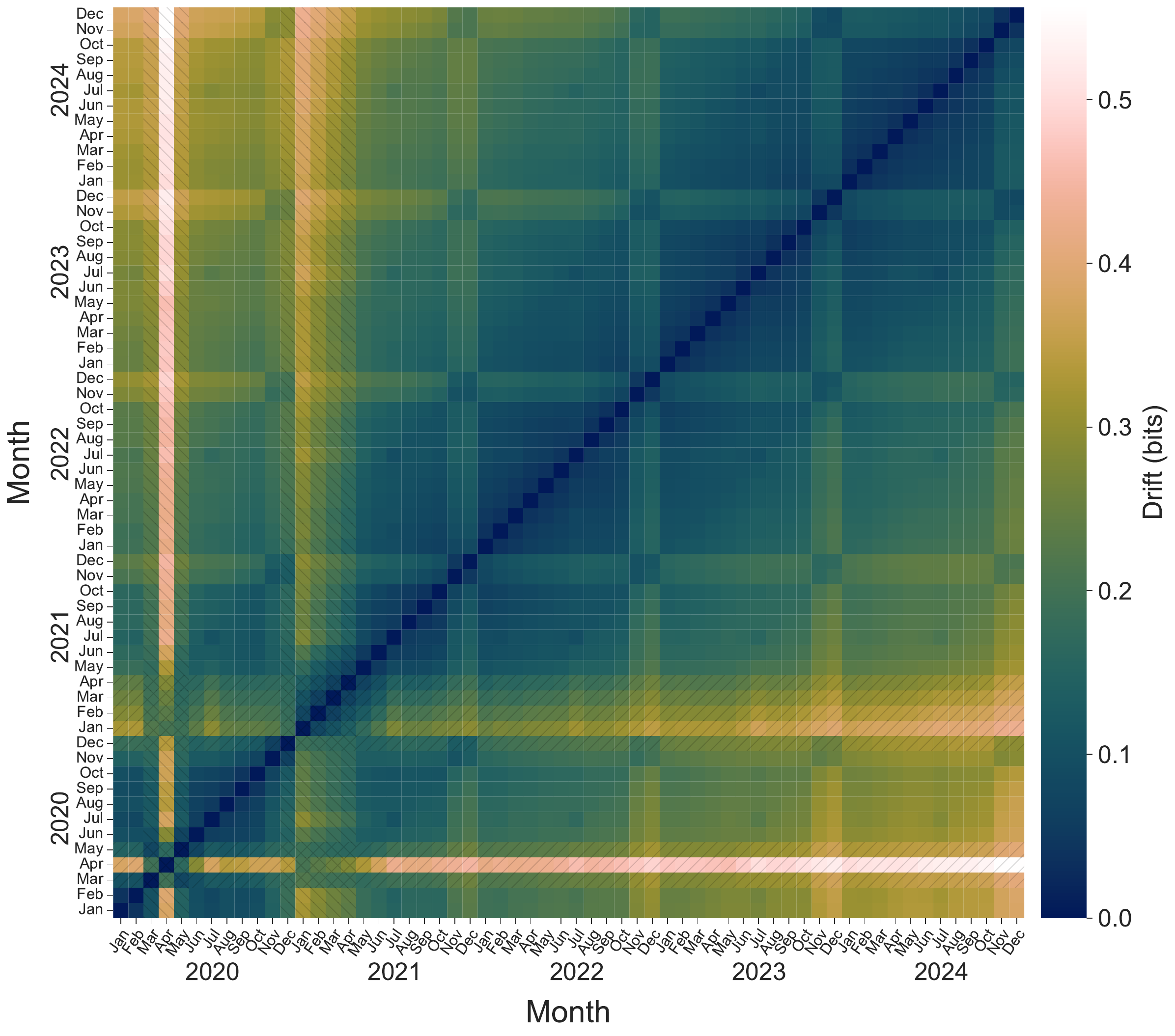}
\caption{
    Drifts in the collective attention for \emph{children's books} between all pairs of months in in the dataset. Tiles are marked with a one-way hatch if one month was affected by a COVID-19 lockdown, and a two-way hatch if both months were affected.
} 
\label{drifts_heatmap_children}
\end{figure*}

\begin{figure*}[p] 
\centering
\rotatebox{90}{ 
    \includegraphics[width=0.9\textheight]{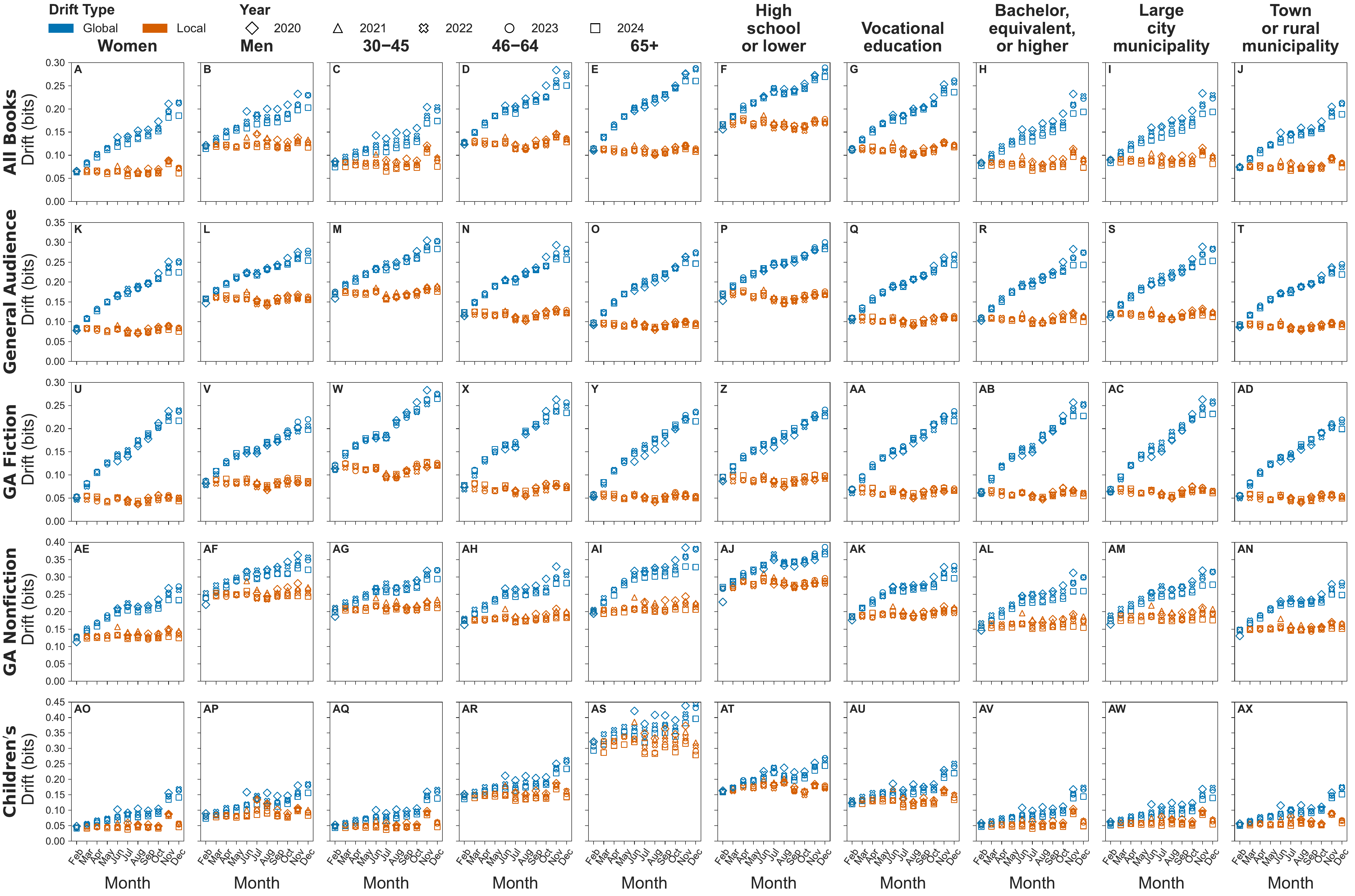}
}
\caption{
    Global and local drifts within each year from 2020 to 2024 for all combinations of book groups and sociodemographic subgroups. Drifts are computed using all loans in each combination of book category and group (not just the top 10,000), which causes some of the drifts to be very high due to random fluctuations.
} 
\label{local_global_subgroups}
\end{figure*}

\begin{figure*}[h!] %[H]
\centering
\includegraphics[width=0.35\textwidth]{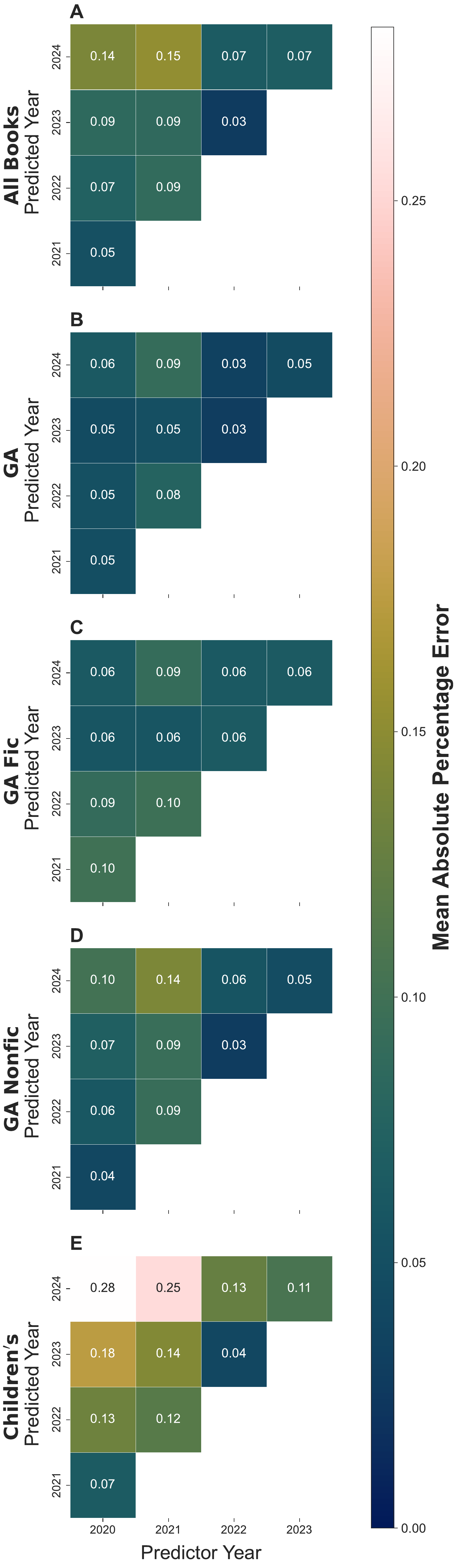}
\caption{
Mean absolute percentage error (MAPE) for predicting year-to-year attention drift at the national level, restricted to local and global drift components. MAPE is computed as the average relative difference between the drift predicted from the preceding year and the observed drift in the subsequent year. Entries involving COVID-affected months are excluded. Lower MAPE values indicate higher predictive accuracy.
}
\label{mape_national_local_global}
\end{figure*}

\begin{figure*}[h!] %[H]
\centering
\includegraphics[width=0.35\textwidth]{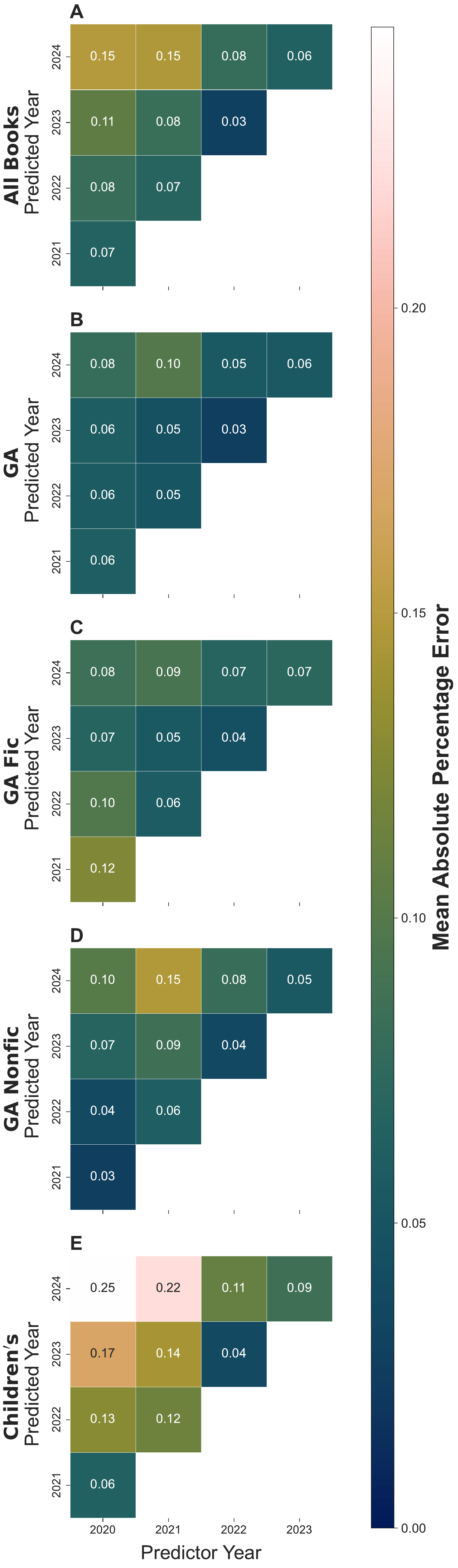}
\caption{
Mean absolute percentage error (MAPE) for predicting year-to-year attention drift at the national level, computed over the full drift matrix excluding the diagonal. Each entry compares the predicted drift from the previous year with the observed drift between the same pair of months. Diagonal entries are excluded because they compare a month with itself and are therefore trivially zero. COVID-affected months are excluded from the calculation of MAPEs.
} 
\label{mape_national_full}
\end{figure*}

\begin{figure*}[p] 
\centering
\rotatebox{90}{ 
    \includegraphics[width=0.85\textheight]{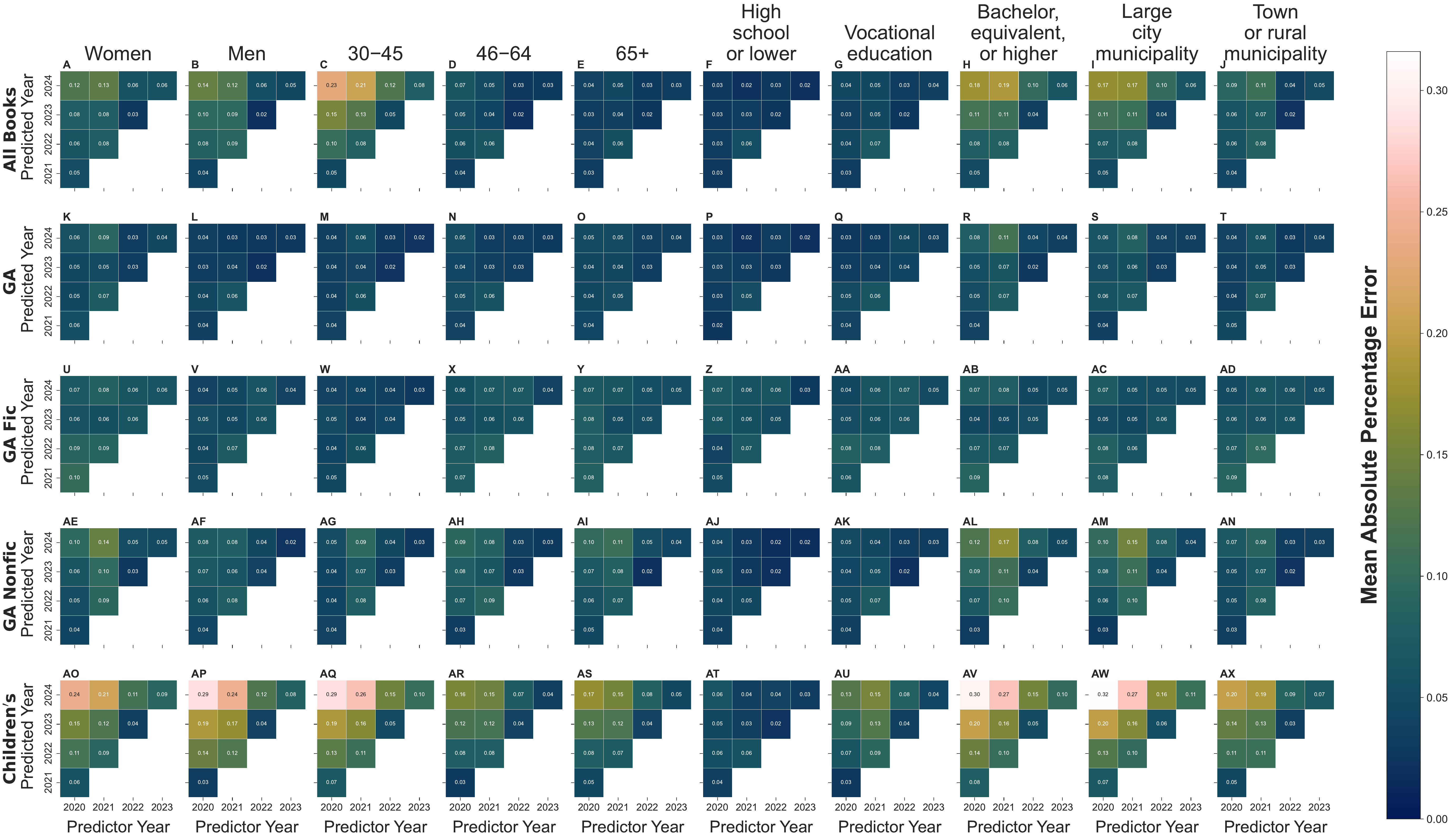}
}
\caption{
Mean absolute percentage error (MAPE) for predicting year-to-year attention drift across loaner subgroups, restricted to local and global drift components. MAPE is computed as the average relative difference between the predicted drift from the preceding year and the observed drift in the subsequent year. COVID-affected months are excluded from the calculation of MAPEs.
} 
\label{mape_subgroups_local_global}
\end{figure*}

\begin{figure*}[p] 
\centering
\rotatebox{90}{ 
    \includegraphics[width=0.85\textheight]{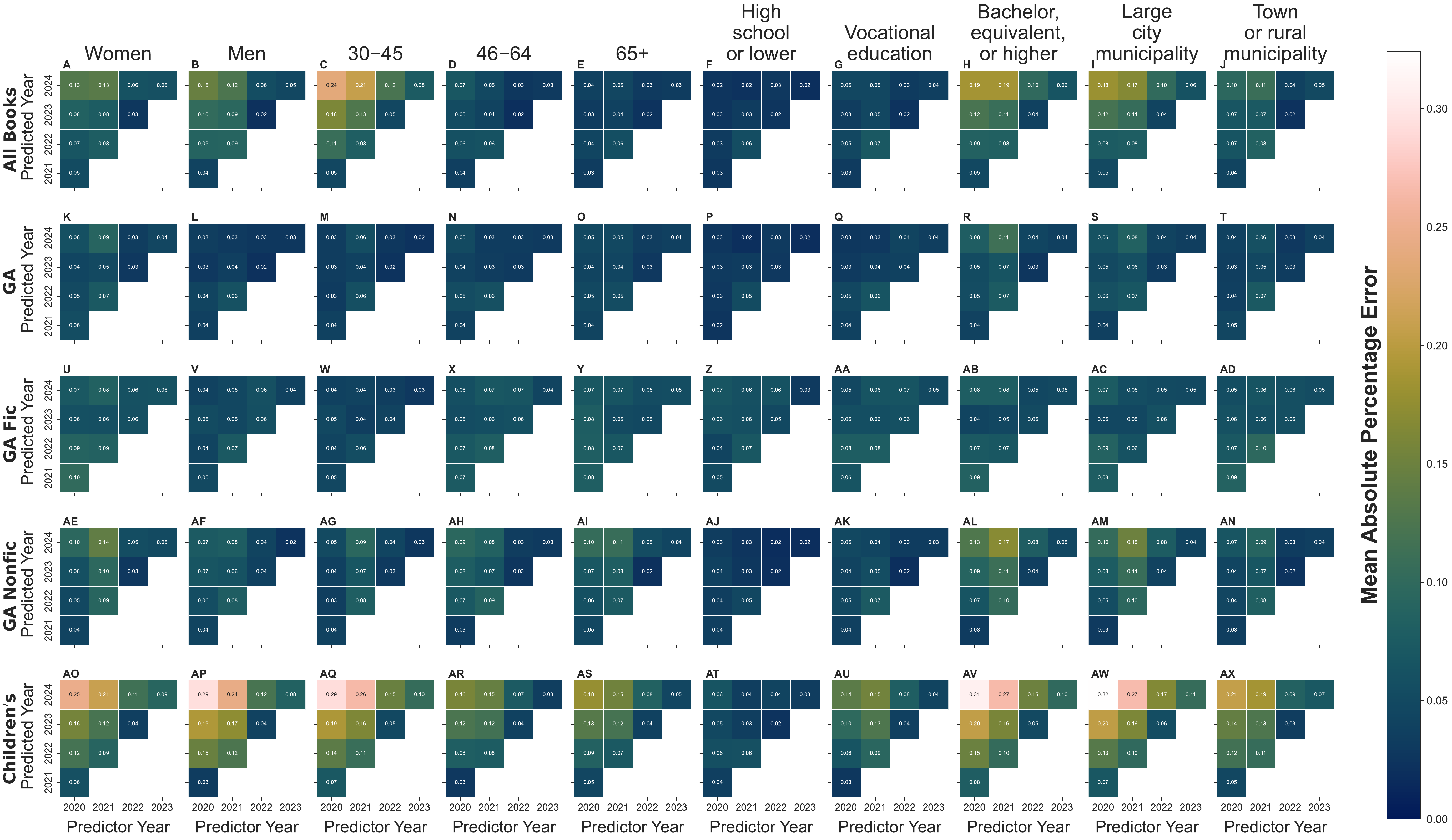}
}
\caption{
Mean absolute percentage error (MAPE) for predicting year-to-year attention drift across loaner subgroups, computed over the full drift matrix excluding the diagonal. Each entry compares the predicted drift from the previous year with the observed drift between the same pair of months. Diagonal entries are excluded because they compare a month with itself and are therefore trivially zero. COVID-affected months are excluded from the calculation of MAPEs.
}
\end{figure*}

\endgroup

\clearpage

\bibliography{sn-bibliography}